\documentclass[12pt,letter]{article}
\usepackage{ifthen} 
\usepackage{booktabs}
\usepackage{multicol}
\usepackage{multirow}
\usepackage{siunitx}
\usepackage[style=ieee, backend=biber]{biblatex}
\newboolean{pdflatex}
\setboolean{pdflatex}{true} 
\newboolean{articletitles}
\setboolean{articletitles}{true} 
\newboolean{uprightparticles}
\setboolean{uprightparticles}{false} 
\def\paperauthors{HEV Collaboration} 
\def\paperasciititle{The HEV Ventilator Proposal} 
\def\papertitle{The HEV Ventilator} 
\def\paperkeywords{{Ventilator},{COVID-19}} %
\def\papercopyright{\the\year\ CERN } 
\def\paperlicence{CC-BY-4.0 licence}
\def\paperlicenceurl{https://creativecommons.org/licenses/by/4.0/}
\usepackage{url}

\usepackage[top=1in, bottom=1.25in, left=1in, right=1in]{geometry}

\usepackage{microtype}
\usepackage{lineno}  
\usepackage{xspace} 
\usepackage{caption} 

\usepackage{subfigure}

\usepackage{graphicx}  
\usepackage{color}
\usepackage{colortbl}
\graphicspath{{./figs/}} 

\usepackage{amsmath} 
\usepackage{amssymb}
\usepackage{amsfonts}
\usepackage{upgreek} 

\newcommand*\patchAmsMathEnvironmentForLineno[1]{%
\expandafter\let\csname old#1\expandafter\endcsname\csname #1\endcsname
\expandafter\let\csname oldend#1\expandafter\endcsname\csname
end#1\endcsname
 \renewenvironment{#1}%
   {\linenomath\csname old#1\endcsname}%
   {\csname oldend#1\endcsname\endlinenomath}%
}
\newcommand*\patchBothAmsMathEnvironmentsForLineno[1]{%
  \patchAmsMathEnvironmentForLineno{#1}%
  \patchAmsMathEnvironmentForLineno{#1*}%
}
\AtBeginDocument{%
\patchBothAmsMathEnvironmentsForLineno{equation}%
\patchBothAmsMathEnvironmentsForLineno{align}%
\patchBothAmsMathEnvironmentsForLineno{flalign}%
\patchBothAmsMathEnvironmentsForLineno{alignat}%
\patchBothAmsMathEnvironmentsForLineno{gather}%
\patchBothAmsMathEnvironmentsForLineno{multline}%
\patchBothAmsMathEnvironmentsForLineno{eqnarray}%
}

\usepackage{hyperxmp}
\usepackage[pdftex,
            pdfauthor={\paperauthors},
            pdftitle={\paperasciititle},
            pdfkeywords={\paperkeywords},
            pdfcopyright={Copyright (C) \papercopyright},
            pdflicenseurl={\paperlicenceurl}]{hyperref}

\usepackage[colorinlistoftodos,textsize=scriptsize]{todonotes}

\usepackage[bottom,flushmargin,hang,multiple]{footmisc}

\usepackage[all]{hypcap} 

\usepackage{cleveref}

\usepackage{longtable} 
\usepackage{siunitx}

\usepackage{xspace}

\def\cmWater {\ensuremath{\,\mathrm{cm\,H_2O}}\xspace}

\def\Pinsp {\ensuremath{\mathrm{p_{inh}}}\xspace}
\def\pcac {\mbox{PC--A/C}\xspace}
\def\pcacprvc {\mbox{PC--A/C--PRVC}\xspace}
\def\valveinhale{\mbox{\tt valve\_inhale}\xspace}

\def\valveoin{\mbox{\tt valve\_O$_2$\_in}\xspace}
\def\valveairin{\mbox{\tt valve\_Air\_in}\xspace}

\def\Ppatient{\mbox{\tt P\_patient}\xspace}
\def\dPpatient{\mbox{\tt dP\_patient}\xspace}

\def\FIOxygen{\ensuremath{\mathrm{FIO_2}}\xspace}

\newcommand{\uC}{Microcontroller\xspace}
\newcommand{\rpi}{Raspberry Pi\xspace}

\begin{document}
\renewcommand{\thefootnote}{\fnsymbol{footnote}}
\setcounter{footnote}{1}

\begin{titlepage}
\pagenumbering{roman}

\vspace*{-1.5cm}
\centerline{\large EUROPEAN ORGANIZATION FOR NUCLEAR RESEARCH (CERN)}
\vspace*{1.5cm}
\noindent
\begin{minipage}{0.6\textwidth}
\includegraphics[width=.25\textwidth]{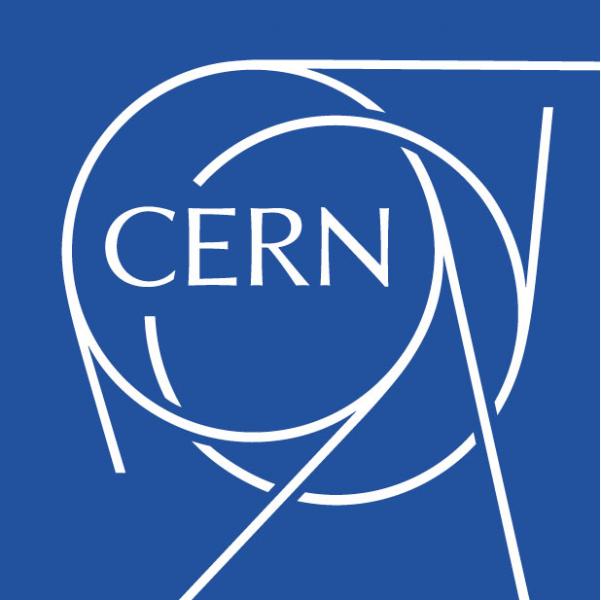}
\includegraphics[width=.4\textwidth]{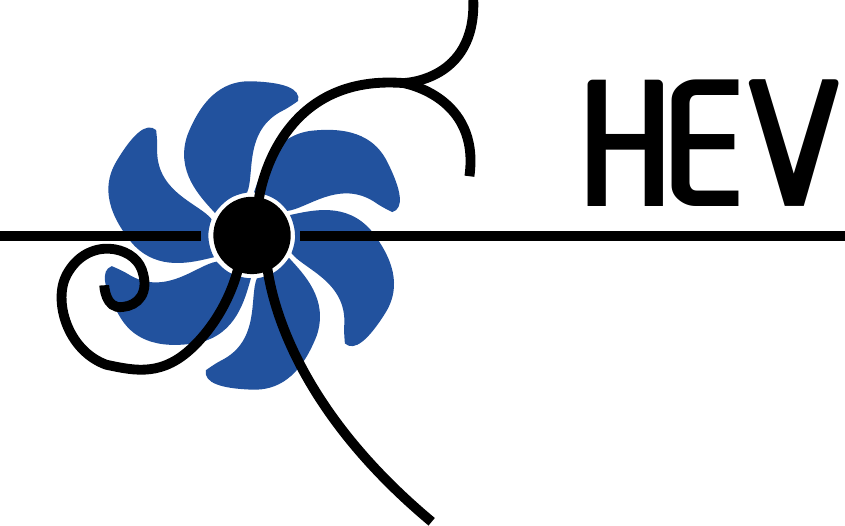}
\end{minipage}
\begin{minipage}{0.4\textwidth}
\begin{flushright}
CERN-EP-TECH-NOTE-2020-002\\
\today 
\end{flushright}
\end{minipage}\\

\vspace*{4.0cm}
{\normalfont\bfseries\boldmath\huge
\begin{center}
  \papertitle 
\end{center}
}
\vspace*{2.0cm}
\begin{center}
\paperauthors{\footnote{Authors are listed at the end of this paper.}}
\end{center}
\vspace{\fill}
\begin{abstract}
  \noindent
  HEV is a low-cost, versatile, high-quality ventilator, which has been designed in response to the COVID-19 pandemic.  The ventilator is intended to be used both in and out of hospital intensive care units, and for both invasive and non-invasive ventilation.  The hardware can be complemented with an external turbine for use in regions where compressed air supplies are not reliably available.  The standard modes provided include \pcac (Pressure Assist Control), \pcacprvc (Pressure Regulated Volume Control), PC-PSV (Pressure Support Ventilation) and CPAP (Continuous Positive Airway Pressure).   HEV is designed to support remote training and post market surveillance via a web interface and data logging to complement the standard touch screen operation, making it suitable for a wide range of geographical deployment.  The HEV design places emphasis on the quality of the pressure curves and the reactivity of the trigger, delivering a global performance which will be applicable to ventilator needs beyond the COVID-19 pandemic.  This article describes the conceptual design and presents the prototype units together with their performance evaluation. 
\end{abstract}
\vspace*{2.0cm}

\begin{center}

\end{center}
\vspace{\fill}
{\footnotesize 
\centerline{\copyright~\papercopyright. \href{\paperlicenceurl}{\paperlicence}.}}
\vspace*{2mm}
\end{titlepage}
\newpage
\setcounter{page}{2}
\mbox{~}
\renewcommand{\thefootnote}{\arabic{footnote}}
\setcounter{footnote}{0}
\tableofcontents
\cleardoublepage
\pagestyle{plain} 
\setcounter{page}{1}
\pagenumbering{arabic}

\section{Motivation}
\label{sec:motivation}

The worldwide medical community currently faces a shortage, especially in low and middle income settings, of medical equipment to address the COVID-19 pandemic~\cite{doi:10.1056/NEJMp2005492,nytimes,doi:10.1056/NEJMp2006141,MALTA2020100384,doi:10.1056/NEJMsb2005114}.  In particular this is the case for ventilators, which are needed during COVID-19 related treatment both in the acute phase, when invasive fully controlled ventilation is needed, and also in the subacute phase during the weaning from mechanical ventilation, which can last for an extended time period.  Companies are scaling up production~\cite{scaling}, but this will not be sufficient to meet the demand according to the current forecasts. There currently exist a wide spectrum of ventilatory support devices, ranging from highly sophisticated through to simpler units~\cite{oxylator,emergency_ventilator}.  
In the context of the COVID-19 pandemic a large number of proposals are already circulating for devices which can be quickly manufactured cheaply and on large scale~\cite{mvm,mvmurl,rice,ventilad,oxvent,belgian,gitlist}.

The pandemic has also drawn attention to the lack of ventilation equipment in low and middle income countries.  Globally, pneumonia is the most common infectious cause of death~\cite{themissingpiece,Zar1052,doi:10.1111/resp.13112}, and the need for adequate respiratory equipment for treatment and management of pneumonia patients will persist even as the COVID-19 pandemic wanes.

In this paper we describe the design of a high-quality ventilator, named HEV (High Energy particle physics Ventilator), which is intended to provide full functionality while being capable of manufacture at relatively low cost, and show the results obtained from the prototype testing.  This is an update from the original proposal~\cite{hev_arxiv1}.  The HEV design~\cite{hev_webpage} is based on readily available and inexpensive components, and the ventilator is intended to be used both in and out of the intensive care unit environment, and for either intubated or non-intubated patients.   The design began in March 2020, using as a starting point the set of MHRA (Medicines and Healthcare products Regulatory Agency) guidelines provided by the UK government regarding Rapidly Manufactured Ventilator Systems~\cite{mhra}.  As the project evolved, regulations and recommendations from other bodies, including the EU, AAMI and WHO~\cite{eu_specs, fda_specs,who_specs} were taken into consideration, the team was reinforced by an international advisory body of clinicians, and advice, collaboration and equipment was provided from local hospitals.   The design and prototypes described here are not, at this moment in time, a medically approved system and will need a process of verification according to medical certification.   However, the design is sufficiently advanced that the functional results can already be shared, as the HEV collaboration moves forward towards certification and manufacture.

\section{Overview of HEV functionality}
\label{sec::mode}

\subsection{HEV operation modes}

Patients affected by COVID-19 face serious issues of lung damage, and the ventilatory equipment must be able to handle situations of rapidly changing lung compliance as well as potential collapse and consolidation. It is critical that the ventilator is able to deliver protective ventilation to patients with nearly normal as well as patients with low compliance.  The driving pressure of the ventilator is a crucial factor for patient outcomes~\cite{hamilton}. In particular, when a low tidal volume is used, the driving pressure is an important variable to monitor and assess the risk of hospital mortality~\cite{amato2015}.  In light of the prolonged recovery/weaning phases involved in COVID-19 critical care cases, there is a need for ventilators which are able to deliver protective controlled ventilation but also are able to delivery assisted ventilation to be efficient for the ventilator weaning process.

Following the COVID-19 guidelines, the HEV development has prioritised pressure modes, and aims to offer these modes in the simplest possible format.    The HEV ventilation modes are divided into two groups: pressure-assist-controlled modes, where the patient effort plays no role or a partial role in the ventilation, and pressure-supported modes, where the patient breathing is spontaneous but supported by the ventilator\footnote{For a pedagogical discussion of ventilator modes, see~\cite{drager}.}.
The former group includes \pcac (Pressure Control -- Assist Control) and \pcacprvc (Pressure Control -- Assist Control -- Pressure Regulated Volume Controlled), while the latter includes PC-PSV (Pressure Control -- Pressure Supported Ventilation),  and in addition provides CPAP (Continuous Positive Airway Pressure).  The CPAP mode is also available, as for lower-cost devices which offer less versatility than the HEV ventilator.  This mode is included with the HEV modes in  order to provide the widest range of support throughout COVID-19 treatment, and may be a crucial option for selection in low resource settings~\cite{cpapimportance}. 

The modes supported by HEV are summarised in table~\ref{tab:vent_mode_table}.   Note that in the interests of simplicity and ease of operation the patient inhalation trigger is to be set for all modes, although the exact trigger levels can be adjusted by the clinician.  This is important to avoid isometric contractions of the diaphragm against a closed valve that could promote diaphragm injuries and prolonged duration of ventilation.  The HEV design also allows PEEP (Positive End-Expiratory Pressure), which is not a ventilation mode in itself but is designed to support steady low positive pressure to the lungs to avoid alveolar collapse. 


\definecolor{columbiablue}{rgb}{0.61, 0.87, 1.0}
\def\blue{\cellcolor{columbiablue}}
\def\bluu{\cellcolor{cyan}}

\begin{table}[h]
\centering
\footnotesize
\begin{tabular}{|p{3.1cm}|p{1.3cm}|p{1.8cm}|p{1.9cm}|p{3.2cm}|p{2.2cm}|}
\hline

\blue \textbf{HEV} 

\textbf{nomenclature}  & \blue \textbf{Patient trigger} & \blue \textbf{Inhalation start}& \blue \textbf{Exhalation start}& \blue \textbf{Respiratory rate (RR)}& \blue \textbf{Comment} \\
\hline 

\blue \textbf{Pressure Control}

\textbf{modes} & \multicolumn{5}{p{8.7cm}|}{} \\
\hline

\bluu \textbf{PC-A/C} & \bluu on & \bluu Machine/ patient triggered & \bluu Machine cycled & \bluu Minimum RR programmed, patient effort can increase RR & \bluu \Pinsp constant \\
\hline

\bluu \textbf{PC-A/C-PRVC} & \bluu on & \bluu Machine/ patient triggered & \bluu Machine cycled & \bluu Minimum RR programmed, patient effort can increase RR & \bluu Volume 

guarantee via \Pinsp variation \\
\hline

\blue \textbf{Pressure Support}

\textbf{modes} & \multicolumn{5}{p{8.7cm}|}{ } \\
\hline

\bluu \textbf{PC-PSV} & \bluu on & \bluu Patient triggered & \bluu Patient cycled & \bluu Spontaneous

In case of apnea fail-safe to \pcac & \bluu \Pinsp constant \\
\hline

\bluu \textbf{CPAP} & \bluu on & \multicolumn{4}{p{10.4cm}|}{\bluu Constant Positive Airway Pressure

  Spontaneous breaths

  In case of apnea fail-safe to \pcac} \\
\hline
\end{tabular}
\caption{Summary of HEV basic ventilation modes.}
\label{tab:vent_mode_table}
\end{table}
\label{subsec::VentiliationModes}

The \pcac mode supplies a defined target pressure to the patient, with a PEEP defined and set by the clinician.  In the case where there is no patient effort, the breathing rate and the inhalation time are fully defined by the parameters entered by the clinician.  The mode also allows the start of inhalation to be triggered by patient effort, detected from the air flow measurement.  The \pcacprvc mode is an extended mode of \pcac ventilation, where the ventilator aims to provide a set tidal volume at the lowest possible airway pressure.  This works by ventilating at an initial set pressure, and if the tidal volume is not achieved, (due to, for example, changes in the patient airway resistance or lung compliance), the ventilation can be gradually adjusted.  In both of these modes the exhalation part of the cycle is triggered by timing, defined by the parameters set by the clinician.

In PC-PSV mode, the inhalation and exhalation parts of the breathing cycle are defined by patient effort and lung mechanics.   The inhalation is triggered as for the \pcac modes, and the exhalation is triggered when the flow drops to a pre-defined fraction of the peak value -- typically 25\%, but can be set by the clinician.

In all modes it is possible to measure the plateau pressure and intrinsic PEEP in order to provide clinical diagnoses and estimation of the patient static lung compliance or to detect AutoPEEP.  The ventilator operation sequence includes a pause time at the end of the inhalation phase during which the valves are closed, normally set to an imperceptibly short time of 5~ms.  The pause time can be increased, for a few breaths, to a few hundred milliseconds, in order to accurately measure to plateau alveolar pressure at zero flow.  The intrinsic PEEP at the end of the exhale phase can be measured in the same way during the pre-inhale state.  These measurements are manual operations performed typically over the course of three or four breaths, and can be used to track clinical parameters.

Particular attention has been paid during the design of HEV to the ability of the machine to synchronise to the patient and to offer suitable pressure profiles, issues which directly influence the patient comfort~\cite{inspiration-comfort}.  This is discussed more completely in section~\ref{sec:testresults} where the HEV prototype measurements are presented.

\subsection{Specifications}
\label{sec::HEVSpecs}

The HEV ventilator is designed to meet the main specifications defined in table~\ref{tab:specifications}.

\newpage

\begin{table}[h]
\begin{tabular}{|p{5.8cm}|p{10.cm}|}
\hline
\textbf{Specification} &  \textbf{Characteristics} \\
\hline
Operation modes & \pcac, \pcacprvc, PC-PSV, CPAP, as defined in the text \\
\hline
PEEP installed & Adjustable in the range  $5-20$\cmWater, in increments of $5$\cmWater \\
\hline
Inhalation airway plateau pressure limit & Set to 35\cmWater by default, with an option to increase to up to 70\cmWater in exceptional circumstances and by positive decision and action by the user \\
\hline
Mechanical fail-safe valve & At 80\cmWater \\
\hline
Mechanical fail-safe valve to connect patient to atmosphere in case of failure   & Present  \\
\hline
Minute volume flow capability & Up to 20 L/min \\
\hline
Inspiratory flow capability & Up to 120 L/min   \\
\hline
Respiratory rate & $10-30$ breaths/min, adjustable in increments of 2 \\
\hline
Inhalation:Exhalation time ratio & 1:2  provided as standard, adjustable in the range 1:1--1:3  \\
\hline
Inhalation time  & Optional setting  \\
\hline
Tidal volume setting & Provided in the range $250-2000$~mL in steps of 50~mL \\
\hline
Inhaled oxygen proportion \FIOxygen & Adjustable between 21\% and 100\% in 10\% steps  \\
\hline
Gas and power supply inlets & Set according to the MHRA standards~\cite{mhra} \\
\hline
Mandatory alarms & Gas or electricity supply failure 

Machine switched off while in mandatory ventilation mode 

Inhalation airway pressure exceeded

Inhalation and PEEP pressure not achieved 

Tidal volume not achieved or exceeded

Hypoventilation and high leakage  \\
\hline
Monitoring of set parameters & Ventilation mode, Respiratory rate, Tidal volume, PEEP, and \FIOxygen \\
\hline
Monitoring of actual measured parameters & Airway pressure, Respiratory rate, Achieved tidal volume, PEEP, \FIOxygen, and real time confirmation of each patient breath in pressure support mode \\
\hline
\end{tabular}
\caption{Summary of the HEV main specifications.}
\label{tab:specifications}
\end{table}

\section{HEV Conceptual Design}
\label{sec::conceptualdesign}

\subsection{Central Pneumatic Unit}
\label{sec::conceptualdesign::pneumatic}

\begin{figure}[ht]
\begin{center}
  \includegraphics[width=0.95\linewidth]{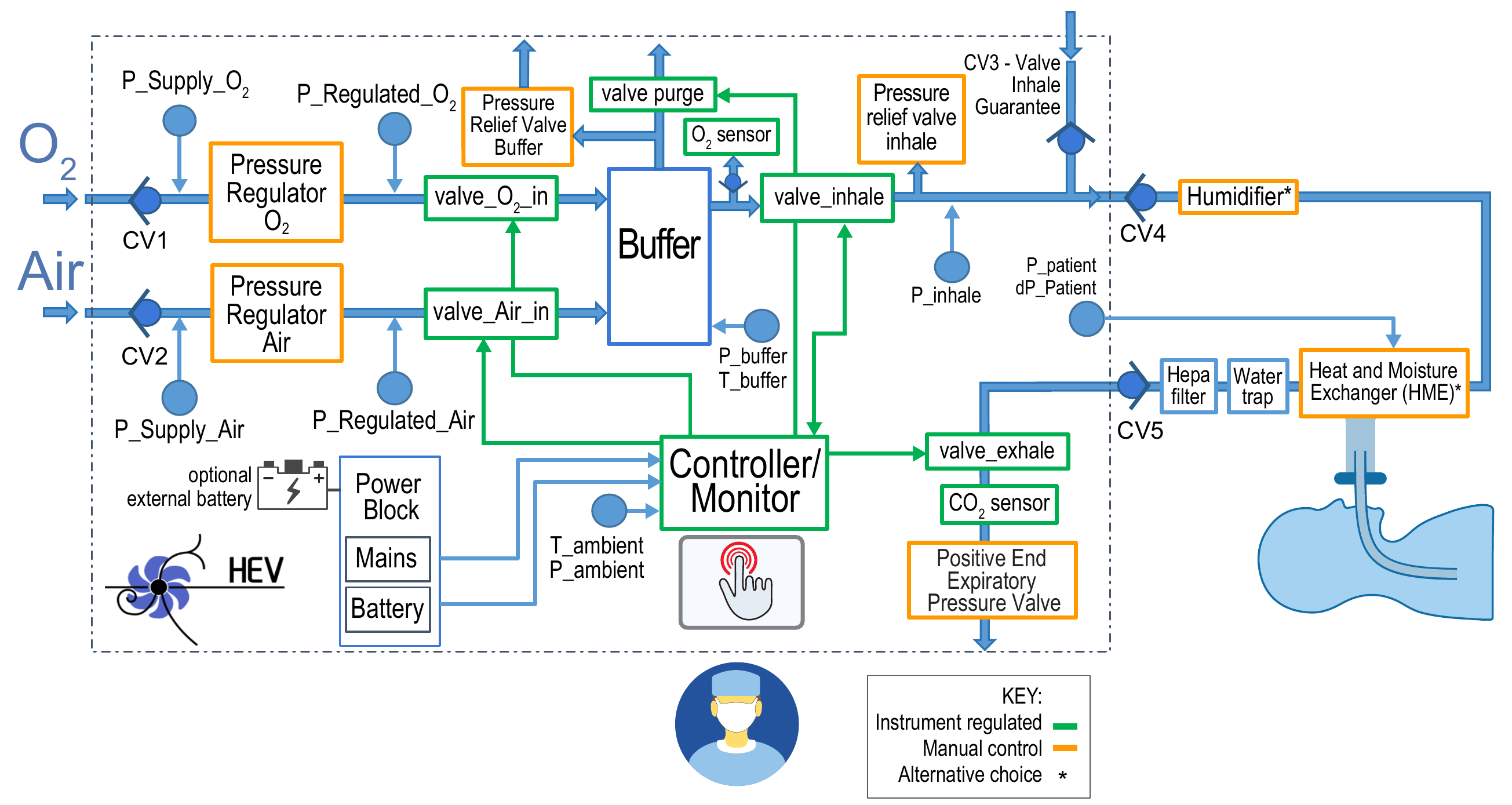}
  \caption{Conceptual design of the HEV ventilator.}
  \label{fig:hev_concept}
  \end{center}
\end{figure}


The HEV conceptual schematic is shown in figure~\ref{fig:hev_concept}. The design is based around a central buffer which pneumatically decouples the ventilator circuit into two, almost independently functioning parts, relating to the filling of the buffer and the gas supply to the patient, respectively.  

\begin{figure}[h]
\begin{center}
  \includegraphics[width=0.3\linewidth]{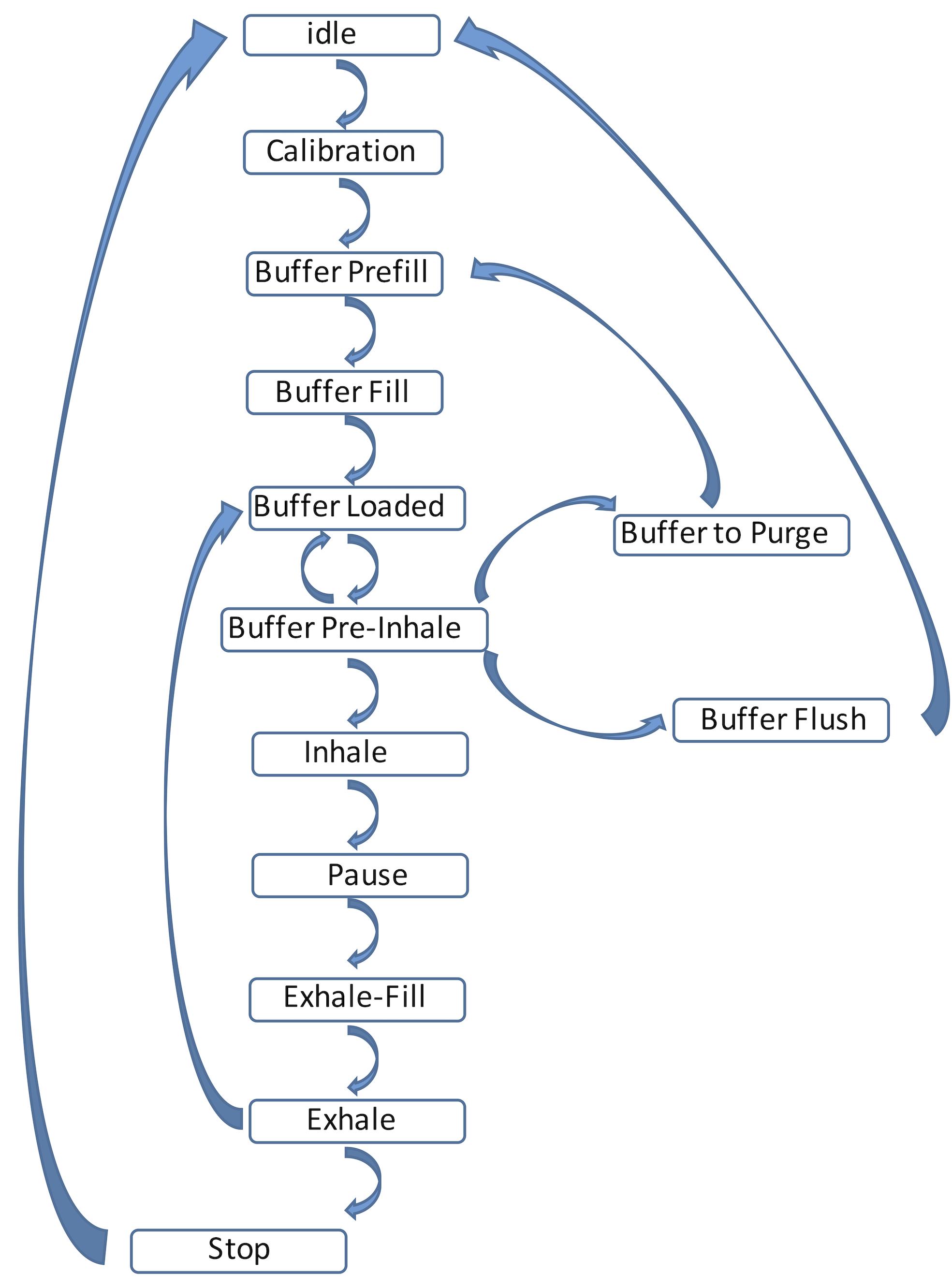}
  \caption{HEV Ventilator State Chart.}
  \label{fig:state_chart}
  \end{center}
\end{figure}

On the filling side of the circuit, two inputs are provided for air and oxygen.  These may be supplied via the standard compressed hospital air and oxygen supplies, in which case typical input pressures will be between 2 and 5 bar.  Alternatively, the input may be supplied via a compressor, oxygen concentrator, or turbine, in which case lower pressures can be expected.  In either case, regulators ensure a step down of pressure to the buffer, and the input valves are used to control the gas flow into the buffer.  The individual flows are passively mixed inside the buffer, which is filled to a target pressure.  Once this has been achieved, the buffer output valve is opened, initiating the respiratory cycle.  This valve is controlled by a PID (Proportional–Integral–Derivative) controller,  which allows a stable delivery of pressure and a fine tuning of the pressure rise time. The controller takes as input the inhale pressure measurement and regulates the valve opening to maintain the inhale pressure at the target value.  This has the advantage that the pressure delivery is independent of flow and buffer pressure.  At the end of the inhalation phase there is a possibility to set a wait time, during which both the inhale and exhale valves are closed.  After the inhalation phase finishes, the exhale valve is opened for exhalation and the buffer is re-filled for the next breath cycle.   The standard state diagram illustrating these actions is shown in figure~\ref{fig:state_chart}.  The state changes are controlled via the microcontroller and may be time or condition driven.   For CPAP operation the input valves to the buffer are kept open and the PID valve is regulated to supply a constant level of pressure.  The PID algorithm ensures that the system is robust against fluctuations in flow or gas supply pressure.

The buffer concept presents many operational advantages.  In general, the separation of the fill and exhale cycle into two separate circuits makes the design, control and component selection more straightforward, and allows less expensive components to be selected.  The initial step-down of the pressure between the supply and the patient introduces safety and robustness against variations in the gas supply.  It also makes the fine tuning of the precise pressure control which is required for the patient side of the circuit more readily accessible.  The buffer volume also avoids that the O$_2$ and air delivery systems need to be able to handle the peak flow rates of up to 120 L/min needed in the inhale phase, and the patient air supply is protected.  The mixing of the gases, which is provided in a natural way inside the buffer, avoids the need to purchase an external gas blender.  In addition the measurement of the O$_2$ concentration, which can be done by spying on the static gas volume, is an inherently more precise measurement than measuring on a gas stream, and does not require a fast reaction time of the meter, nor a medically compatible meter.  Should the design need to be adapted to a more extreme (very hot, or very cold) environment, thermal control of the gas in the buffer is straightforward.  In addition there is a monitoring advantage: the delivered tidal volume can be calculated from the pressure drops in the buffer.  This provides a precious monitoring cross-check in addition to the standard tidal volume measurement, reinforcing the safety of the design.

\begin{figure}[h]
\begin{center}
  \includegraphics[width=0.95\linewidth]{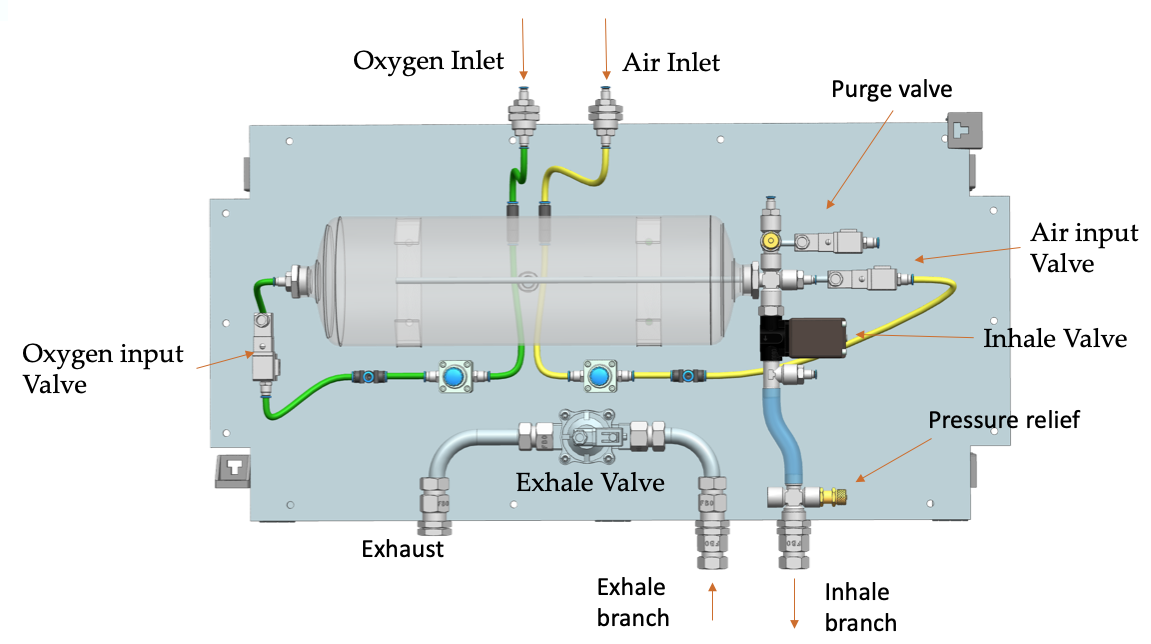}
  \caption{Schematic of the pneumatic part of the HEV system.}
  \label{fig:pneumatic_schematic}
  \end{center}
\end{figure}

To ensure better oxygen intake and to prevent lung collapse, the exhalation branch is fitted with a PEEP valve.
The respiratory rate, inhalation time and pause time are all controllable. If a PEEP pressure is set, then the pressure in the lungs will not fall below the minimum of the PEEP pressure.  For normal vent mode operation the pause time is set to be imperceptibly short ($\sim5$ ms), however longer values can be set, for instance when the care giver wishes to measure the lung static compliance. 
The design concept provides inherent safety in that there is a step down of pressure from the regulator before the air reaches the buffer, and in addition the inhale valve is only opened once the pressure is stabilised.  Via the buffer step the system is also inherently protected from variations in the inlet pressures.
The volume of air taken by the patient (tidal volume) can be calculated knowing the buffer pressure measurements, the PEEP value, and the fixed volumes of the buffer and tubes in the system.  The volumes of the tubes, typically between 1 and 2 L, including any equipment such as humidifying devices, may be  routinely confirmed, along with checking for leaks and measuring tube compliance, before the ventilator is connected to the patient. 

The patient is directly protected from over-pressure via the pressure relief valve, which opens at 50\cmWater.  In addition, the pressure  sensor in the buffer continuously monitors for over-pressure. In case of over-pressure in the buffer, the electro-valve is opened to purge its contents and refill it to the correct level.  In case of failure of the output valve of the buffer or a power cut, the system switches to exhalation mode, and the patient is connected to atmosphere via the inhale guarantee valve, which opens mechanically.

A schematic of the central pneumatic part of the system is shown in figure~\ref{fig:pneumatic_schematic}.  This diagram highlights how pure oxygen and air are injected separately into the buffer by using two different valves.
To achieve a good mix, a tube is inserted into the tank from the air side, taking air to be injected close to the oxygen injection point. 
Due to the high flow and the transition of the tube into the larger buffer a turbulent flow is generated, which mixes the oxygen and air. 
The mixture then travels to the other side of the buffer where it is removed via the proportional valve. The concentration of the oxygen is measured and in case it is not correct it is possible to vent the buffer and the cycle can start over. The oxygen levels can be mixed from 21 to 100 percent.  The oxygen fraction is measured with zirconia sensors inserted after the oxygen tank which are reliable, inexpensive and readily available.   The sensors are installed after a bypass, which enables them to be independent of the patient supply and to use a low flow of gas.  As  an additional option, a CO$_2$ sensor can be installed at the exit of the of the respirator and read out from the controller.  CO$_2$ sensors based on non-dispersive (NDIR) technology are inexpensive, have good selectivity, and are no affected by the oxygen level.


\subsection{Alternative air supply}
\label{sec::turbine}

The HEV collaboration is actively investigating alternatives to the compressed air supply available in hospitals.  The major requirement is that the system should be able to fill the HEV buffer in less than one second. 

\begin{figure}[h]
\begin{center}
  \includegraphics[width=0.3\linewidth]{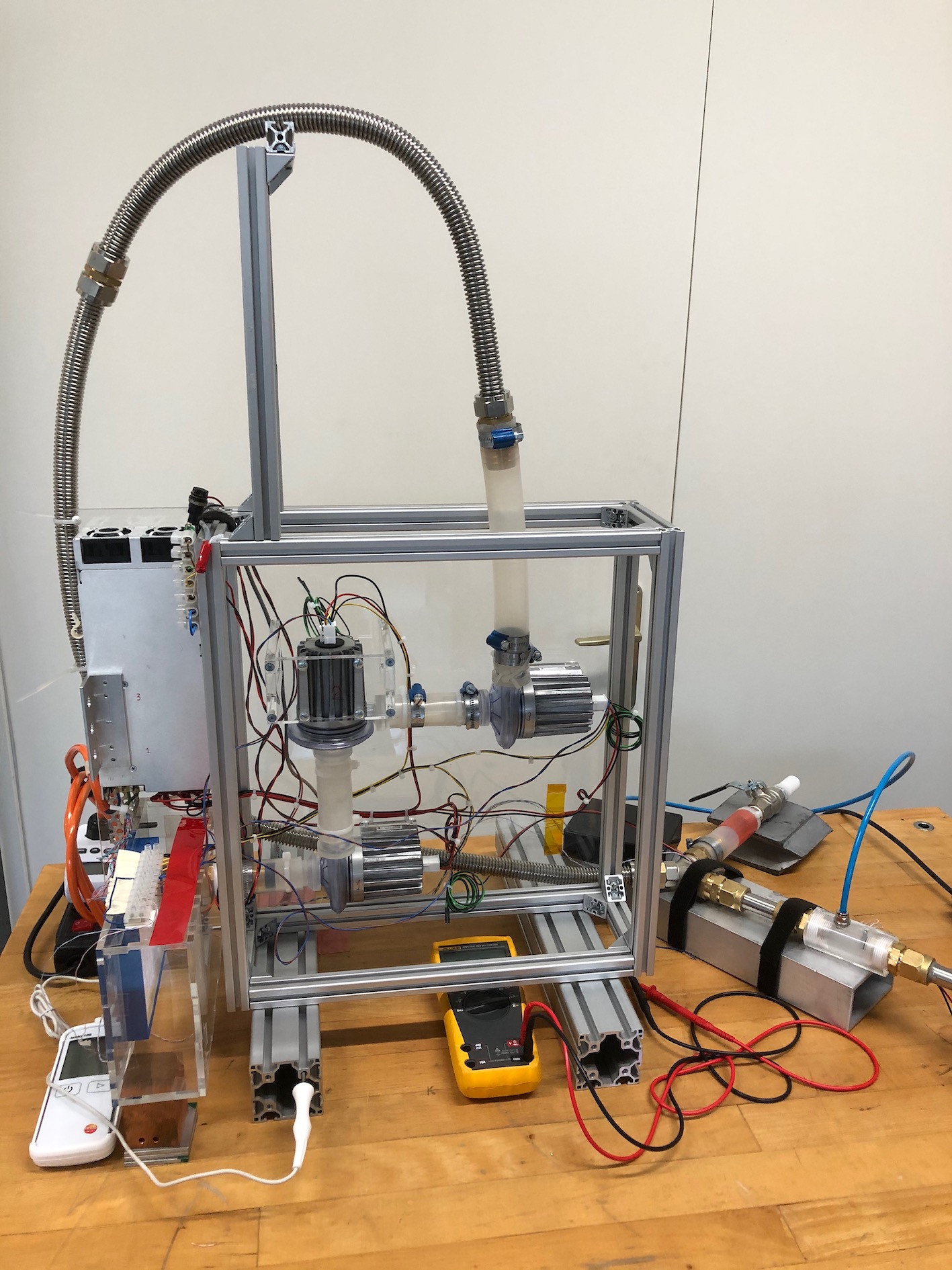}
  \includegraphics[width=0.5\linewidth]{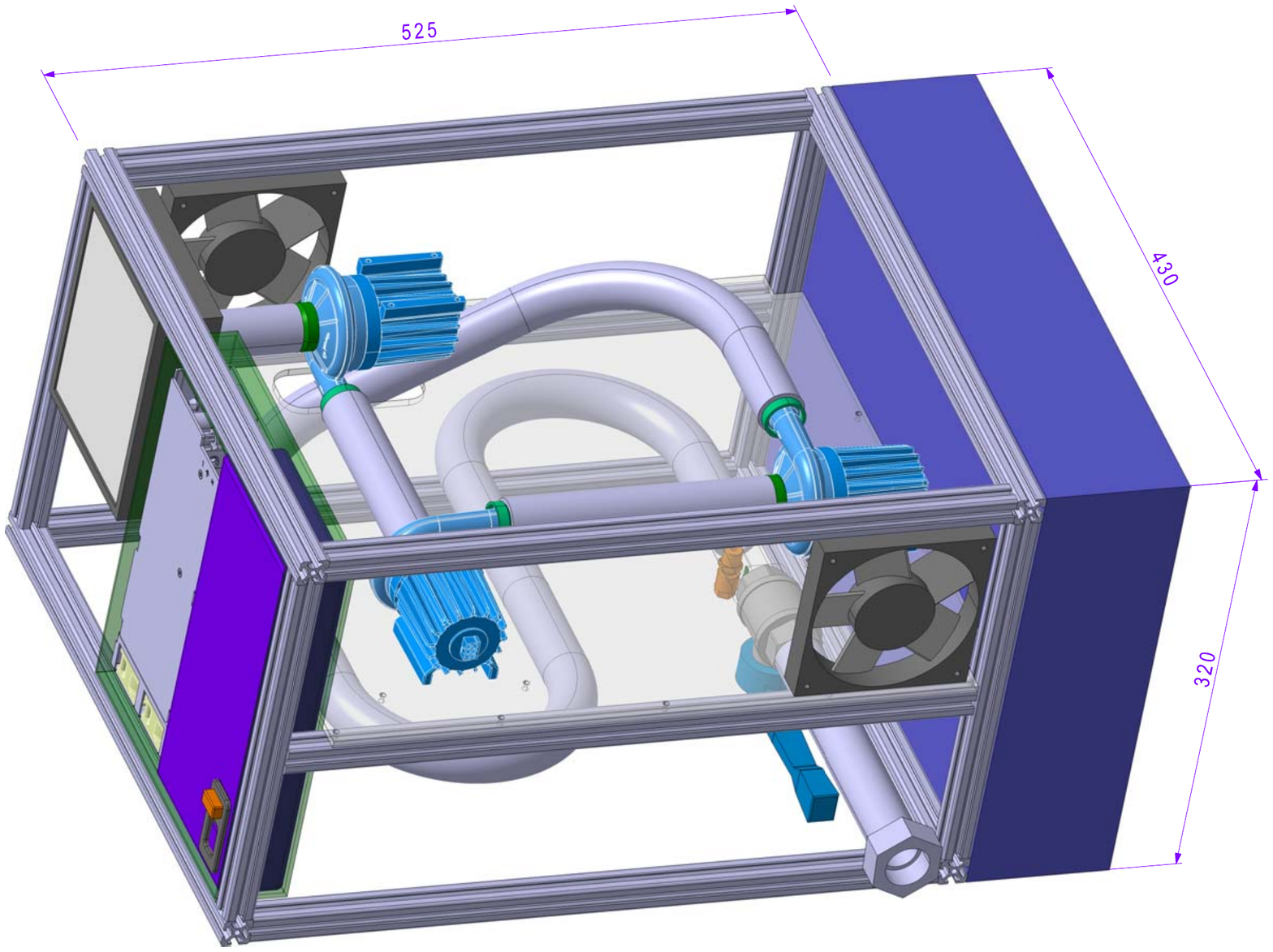}
  \caption{Left: prototype of the turbine system proposed as an alternative to the compressed air supply.
    Right: drawing of the concept for a complete system which is a box divided in two parts:
    the top contains the turbines, the bottom contains the corrugated steel pipe, the thermal and pressure
    sensors and the outlet connector. 
    Mounted on the left side of the box are the air filter, the power supply and the box containing the controller.
    The blue box on the right provides an enclosure for the optional battery system.
    }
  \label{fig:hevturbine}
  \end{center}
\end{figure}

The first option considered is a relatively small and transportable system, based on turbine blowers. A prototype has been built with 3 small turbines in series, powered by 24~V. The speed is controlled by a 0 to 5 voltage level. The prototype is illustrated in figure~\ref{fig:hevturbine}.
The air is filtered by a HEPA (high-efficiency particulate air) filter.
The different parts are interconnected by pieces of 1 inch plastic pipe.
Before reaching the outlet, the air is transported by a corrugated steel pipe, about 1 meter long, acting as an intermediate reservoir. In addition, the corrugated steel pipe allows thermal exchange with the environment, reducing the temperature of the air which is heated in the turbines.  
In the system, the temperatures of the turbines and of the compressed air are continuously monitored with thermal sensors. Pressure sensors are used to maintain the required air flow by acting on the turbine speed. The prototype has been tested by connecting to a $\sim$10~L buffer acting as a HEV simulator, with one input valve and one output valve alternatively working on a 3 to 4 seconds cycle. The test shows that the turbines are able to fill the buffer in the required time and are thermally stable. The change of the temperature in the reservoir was less than \SI{0.5}{\celsius}, after one hour of continuous work.

For increased independence from the hospital setting,
it would be ideal to include an external battery which could power the turbine and the HEV.
Suitable candidates, of the order of 80~Ah, 24~V can be found on the market, which also include the option to be recharged by solar panel.
A relatively low-cost option which will be tested in the next prototype
is based on a battery from the e-bike market.
Figure~\ref{fig:hevturbine}, right, shows the concept for a complete turbine-based system. The system is divided in two parts separated by the support for the turbines. The corrugated steel pipe, temperature and pressure sensors, and the outlet connector are at placed behind the support. The air filter, power supply, and a box with a controller are on one side of the structure. An optional battery system can be installed at the other
side.  A system of low-power fans is used to cool the turbines. 

An alternative to a custom designed turbine unit would be a commercial oil-free compressor. Suitable candidates have been identified with high availability and low cost and are being tested in parallel with the turbine system.

\subsection{Accessories and disinfection protocols}
\label{sec::turbineaccessories}

The choice of breathing circuit is an element of the system which is considered as an accessory which can be finalised in the manufacturing stage of the ventilator.  
HEV has been tested with concentric tube geometry and a double limb circuit, and could be supplied with an adapter to use either method.  

The breathing circuit can also be equipped with humidifier or a heat and moisture exchanger (HME) filter by choice, which is necessary to protect the patient, who may be dehydrated, from the dry medical or ambient air.  
The additional volume and the small pressure drop introduced by the humidifier device will be taken into account during the calibration setup phase of HEV and do not affect the operational parameters in any way. 
Alternatively, an HME filter might be used to maintain humidity levels and provide particle filtering, particularly if there is a worry that condensation in the flow pressure sample lines can affect the monitoring.  
The HEV will likely feature a HEPA filter to be used before the exhale valve to filter 99.9\% of bacteria and virus, allowing the exhale to be released to atmosphere.  
In this case there would not be a need for a scavenger circuit. 

All equipment that comes in contact with the patient needs to be either changed or disinfected and sterilised after every patient. 
There are multiple options to do this and currently autoclave cleaning is supported, 
%
i.e all the materials which will be reused need to withstand a temperature of $132^\circ$C for up to 4 minutes. 
The entire exhaust block may be easily dismounted and swapped with a spare block, so that the ventilator can continue to be used for the next patient while the block undergoes steam or autoclave sterilisation.

The PEEP valve is a purchased accessory. 
Care must be taken to select a valve with sufficient range and precision and priority will be given to reusable components which can be cleaned by chemical disinfection or autoclave procedures. 

\subsection{Electrical design}
\label{subsec::electricaldesign}

 The electrical design has focused on rapid production for the HEV prototype.  It has been implemented as two parts to allow flexibility in design choices and to simplify future modifications: a motherboard for hosting a microcontroller, interfacing to valves, sensors, and connecting to the user interface provided by a \rpi via a touchscreen; and a power system, including an uninterruptible power supply (UPS), to provide the power to the motherboard and components connected to it.
 
The motherboard is the core of the system with the following features. 
A connector serves to mount and power the \rpi and access the signals on its general-purpose-input-output (GPIO) bus.  The board also has connectors to mount and power the ESP32 microcontroller, and access the input/output signals for valve control, sensing and other ancillary functions. There are integrated circuits to drive control signals to valves and the corresponding connectors for cabling to the valves. These circuits are driven by the ESP32 using pulse-width-modulation appropriate to the choice of valve. Further connectors serve for the cabling carrying signals from sensors. The signals are attenuated on the motherboard before connection to the ESP32 for digitisation. The motherboard also provides a number of embedded sensors for monitoring and debugging, light-emitting-diodes (LEDs) and a buzzer for user information and alarms, 
spare channels for additional valve and sensor connections, and
connectors to allow the powering of fans and the touchscreen.
 
\begin{figure}
\begin{center} 
  \includegraphics[width=0.9\linewidth]{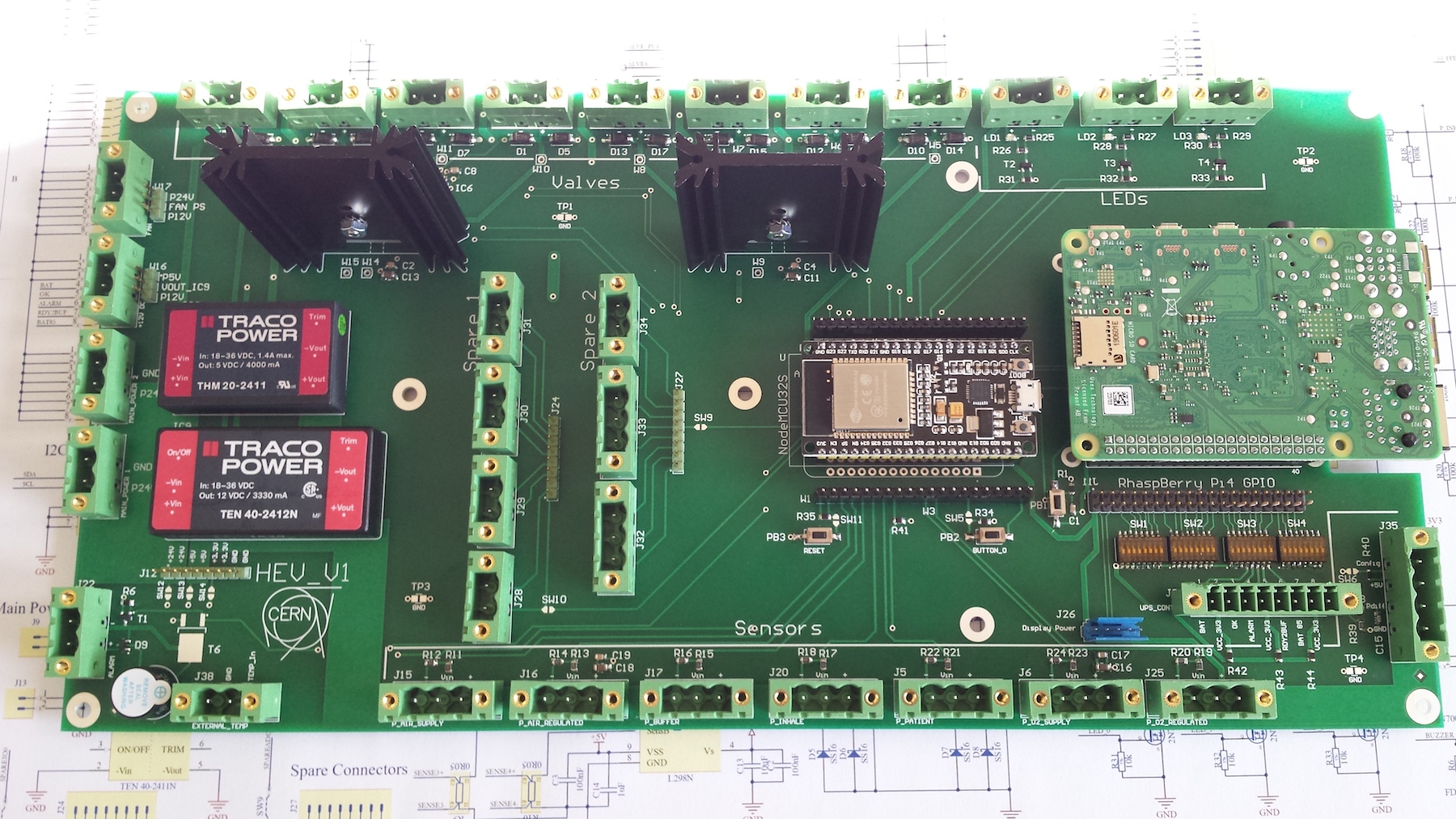}
  \caption{Photograph of the PCB motherboard as implemented for the HEV prototype.}
  \label{fig:hev_PCB}
  \end{center}
\end{figure}

The motherboard (figure~\ref{fig:hev_PCB}) has dimensions $ 322\,{\rm mm} \times 160\,{\rm mm} \times 60\,{\rm mm}$, and is compatible with the mechanical design of the prototype HEV system. It runs on a nominal 24\,V~DC ($18-36$\,V~DC) power supply, and contains local power regulation to generate the lower voltages required for the hosted components.

The power system is modular and uses commercially available components mountable on standard DIN-rails.  The main stage is a stabilised power supply unit with battery management. This generates the 24\,V~DC for the motherboard and is compatible with $110-230$\,VAC.  It is connected with a UPS which will take over if mains power is lost and allows autonomous use of the system for more than 20 minutes.  Configuration of the battery management is done by the \rpi through the motherboard via a specific connector and cable to the UPS.  Connection to the mains power is with an IEC (International Electrotechnical Commission) inlet filter and power is turned on and off with a rotary switch. The DIN-rail mounting is compatible with the mechanical design of the prototype HEV system.

\subsection{Control and User Interface}
\label{subsec::controlanduserinterface}

\subsubsection{Control}

\begin{figure}[ht]
\begin{center}
  \includegraphics[width=0.85\linewidth]{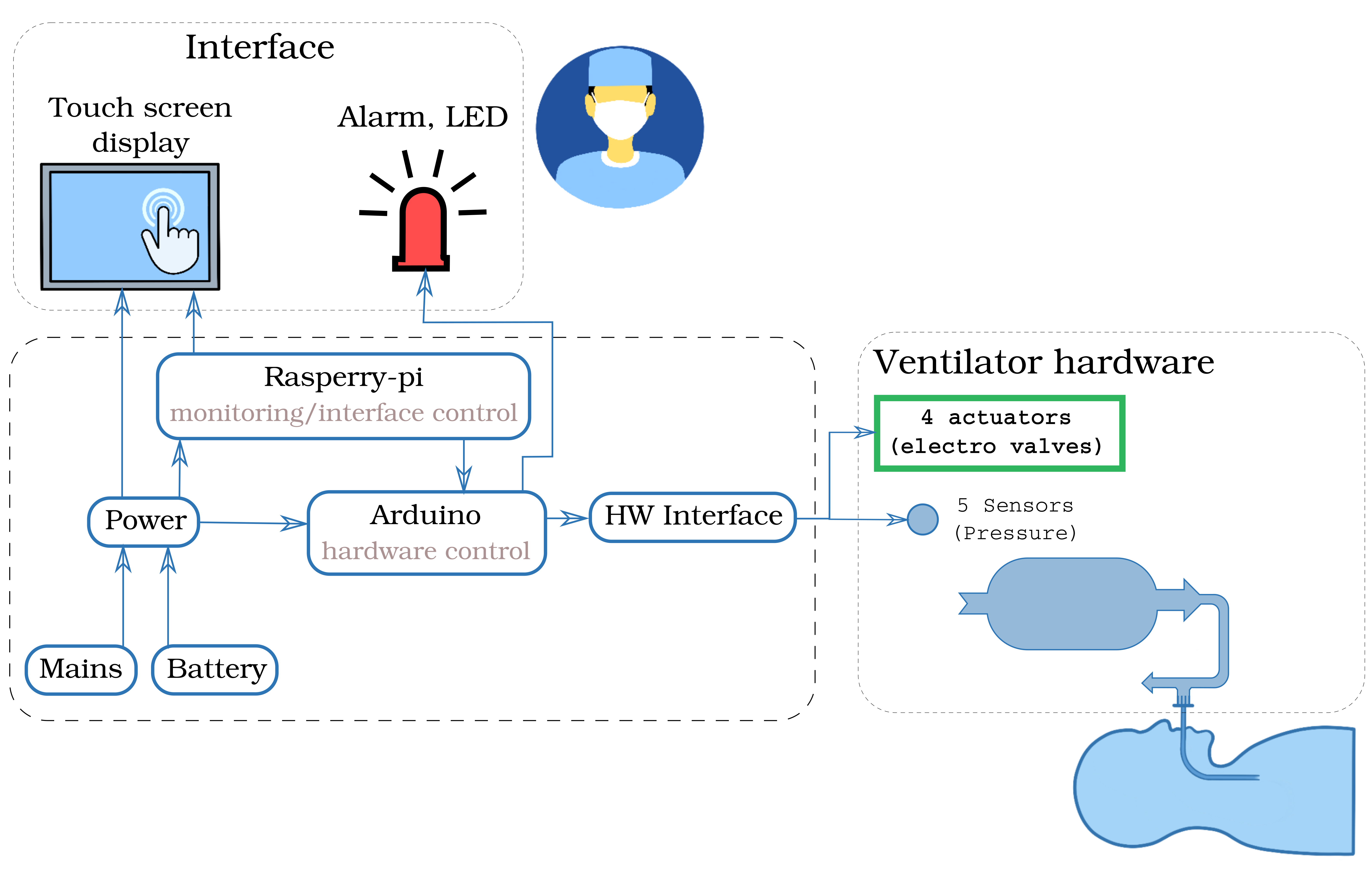}
  \caption{Conceptual layout of the controls and user interface.}
  \label{fig:control_diagram}
  \end{center}
\end{figure}

The control concept, based around the embedded controller receiving the signals from the sensors and valves, and the touchscreen interface to the clinician, is illustrated in figure~\ref{fig:control_diagram}. The control software is implemented directly on the embedded controller, which fully controls the ventilator operation. The controller uses feedback from sensors to control valves in order to achieve the desired pressures.

\begin{figure}
\begin{center}
  \includegraphics[width=\linewidth]{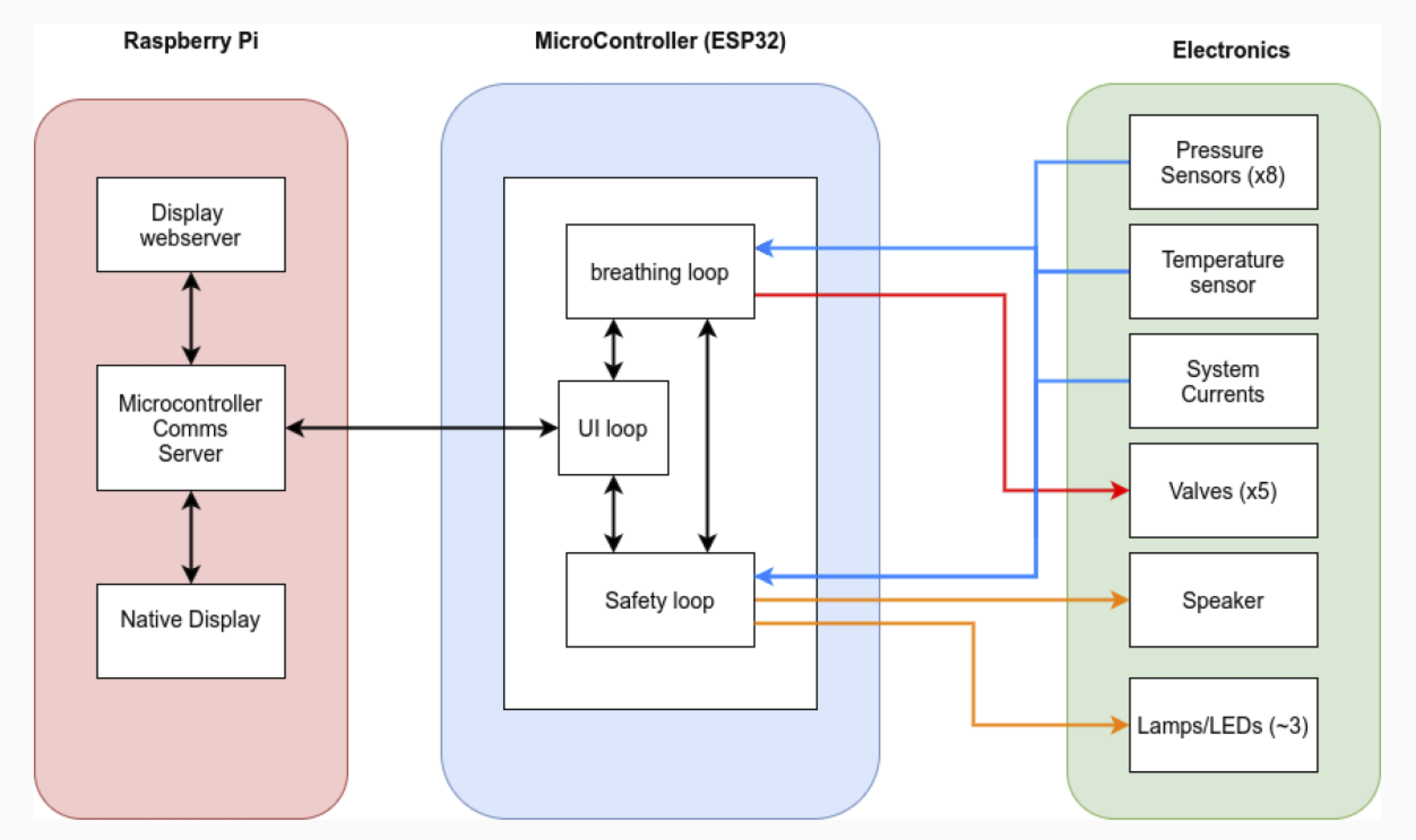}
  \caption[Diagram of the software and computers]{Diagram of the location of the different elements of hardware and software in the two computers (\uC and \rpi) and the circuit board attached to the \uC.}
  \label{fig:hev-sw}
  \end{center}
\end{figure}

The general overview of the control system and the user interface is shown in figure~\ref{fig:hev-sw}.  Central to the operation and monitoring is the \uC. 
An ESP32 \uC chip has been chosen at the time of prototyping, due to its high availability and low cost.   This is a dual core, single process Arduino compatible controller with additional CPU power and memory.  It runs with no operating system which can be an advantage from the point of view of reducing complexity and promoting stability.  Several alternatives exist and could be used depending on local availability in different geographical locations.
All of the primary functions corresponding to the breathing function of the patient are controlled by the \uC.   The \uC connects to the electronics to read the pressures, flows and temperatures.  It controls the opening and closing of the valves and provides a simple status/warning via a speaker and a set of LEDs.    Interconnect electronics is provided via the intermediate PCB (not shown) which steps up and down voltages where necessary.  

The \uC is connected to a \rpi which handles communications to the touch screen, WiFi and Ethernet, and controls the displays.  If the communications are interrupted, the ventilator continues to run normally.  A web server is also provided such that display information can be seen remotely (although this is not considered an essential function for operation).

\subsubsection{\uC functionality}

The \uC software design consists of three distinct threads of processing: 
\begin{itemize}
    \item  A breathing loop responsible for operating the valves in a manner corresponding to the ventilator modes, such as \pcac or PC-PSV, as described in section \ref{subsec::VentiliationModes}.  
    \item A User Interface (UI) loop for relaying the current status and readings to the \rpi, and for accepting commands for setting modes and parameters.
    \item A Safety Loop which is responsible for raising alarms when patient or system readings deviate from acceptable limits.
\end{itemize}

For all loops a finite state machine is employed to ensure operation is achieved in a repeatable and deterministic manner.   The state machines operate semi-independently, with defined interfaces for information to pass between them.  For example, this allows the Safety Loop to know both the current parameters set by the UI loop, and the current breathing loop ventilation mode.  
This allows each state machine to be tested individually by expressly checking the state response to each input at the interface.

The \uC design allows the breathing and safety loops to continue to operate in the event of loss of connection to the \rpi via the UI loop.
Visual and audible alarms signify a loss of connection but breathing and safety functions continue whilst the connection is being re-established. 
The lack of an underlying operating system in the \uC reduces the probability of system lock-up/failure caused by other processes running in the background.

\subsubsection{\rpi functionality}
The \rpi software is designed to display the readings from the \uC in a UI on the touchscreen.  
Both the webserver and the native displays receive the same information.  Both can be used to send control messages to the \uC.  
However, in the case of the webserver, the controls may be disabled for reasons of security and safety.  Both interfaces employ the same layout, such that they appear as similar as possible. 
Following the advice 
of clinicians, the user interface has been designed based on the following concepts:
\begin{itemize}
    \item Clear text, symbols and graphs which can be seen from the end of a hospital bed through PPE.
    \item Neutral colours when all is OK, and by contrast, flashing indicators and messages when there are alarms.  In general, safe and unsafe status should be easily distinguished. 
    \item Screen locking/unlocking feature (with a timeout) to prevent accidental touchscreen presses.
    \item Confirmation of all parameter changes  (two parameter changes would require two separate confirmations).
    \item Setting and reading back information (so that they can be compared).
    \item Simple navigation: no setting/parameter should be more than two clicks away.  Normal operation should be separated from calibration/expert testing, to maintain an uncluttered interface.
    \item Interface should be touchscreen friendly: items should be placed far enough apart to minimise accidental mis-clicks.
    \item Familiarity: designed to look familiar to clinicians, similar to already existing ventilator interfaces.
\end{itemize}

\subsubsection{Communication}

Communication between the \uC and the \rpi is achieved using the High-level Data Link Control Protocol (HDLC) [ISO 13239:2002].  
A subset of the full protocol functionality was implemented, including supervisory (ACK/NACK frames) and  information (INFO) frames.  
Information frames are subdivided into Data, Command and Alarm subtypes.  
Error detection is provided by a checksum (CRC16-CITT).  
More details are  available in \cite{sw-protocol}.  The protocol implementation is provided as a set of libraries on both the \uC and the \rpi.

On the \rpi the protocol library is employed by a dataserver which broadcasts the data and alarm messages from the \uC to the  user interface processes.  Similarly, Command messages are relayed from the user interfaces in the reverse direction.  This is the expected method of operation, however, any message type can be sent in either direction. JSON is used as the communication format for messages from the dataserver to the user interfaces. These messages are built from the data, command, and alarms formats defined in the HDLC protocol library such that duplicate definitions are not required in JSON. More detail on the dataserver is given in \cite{sw-dataserver}.

\subsubsection{Software Practice}

The software development has been done in a robust and flexible manner, prioritising considerations of component failure from the start.  The development team has been organised into pairs of developers working in the same domains, with strong familiarity with the code such that no individual is irreplaceable.  Readability and simplicity of code has been favoured over complexity, for ease of bug tracing and testing.  For the prototype, the guidelines of IEC 62304 have been used, and internal checking/assertions are included in the software.  The communication protocols follow the High Level Data Link Control recommendations from ISO/IEC 13239:2002.  In particular, data transmission is done with acknowledgement, and checksums are performed to confirm the data integrity.  While the software system in place for the prototype is not yet fully qualified to ISO standards, the structure has been set up in such a way as to show that this is possible.

\subsubsection{User Interface}
\label{sec::UI}

\begin{figure}[h]
     \includegraphics[width=\textwidth]{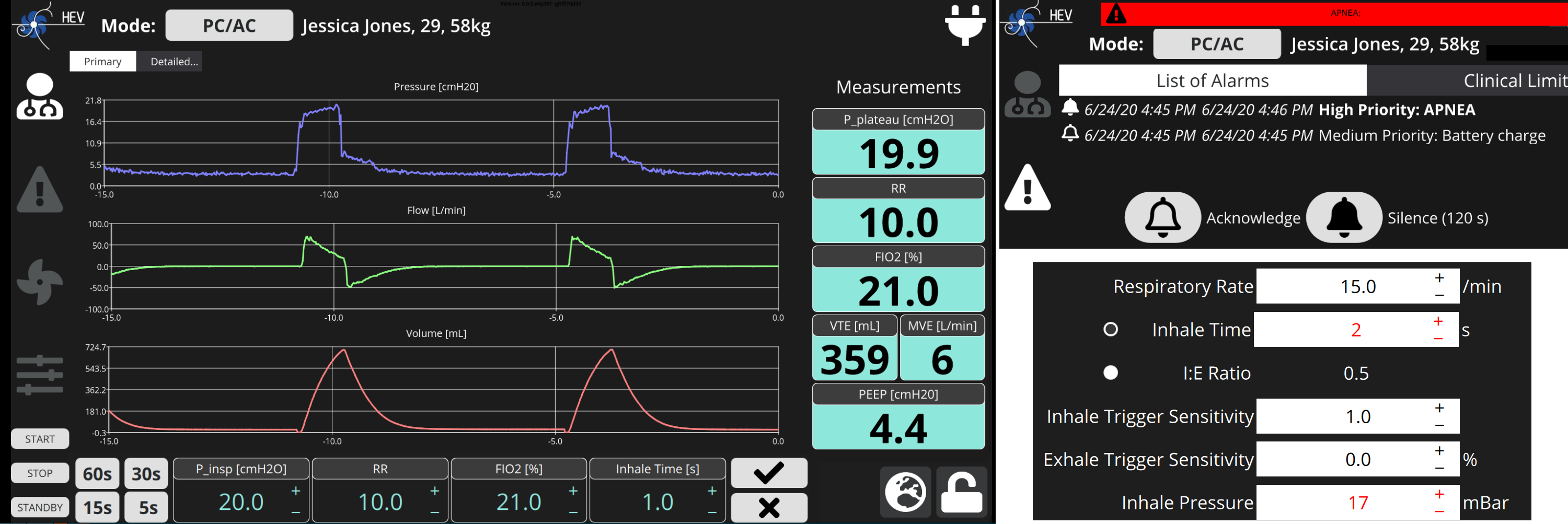} 
    \caption{Example displays of the Native User Interface.  Numbers and graphs are purely indicative.}
    \label{fig:nativeui-homepage}
\end{figure}

\begin{figure}[h]
    \centering
    \includegraphics[width=0.45\textwidth]{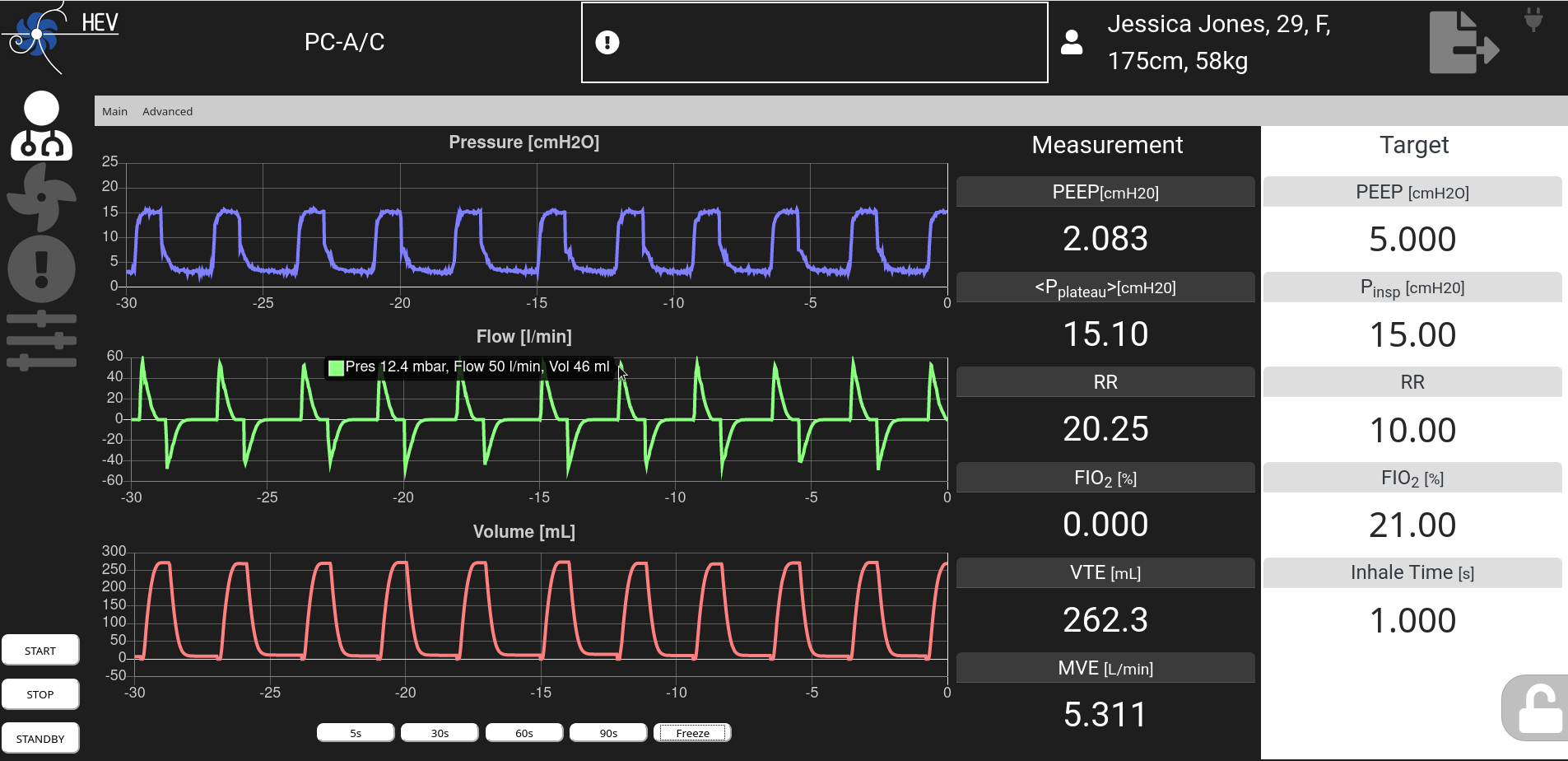}
    \includegraphics[width=0.5\textwidth]{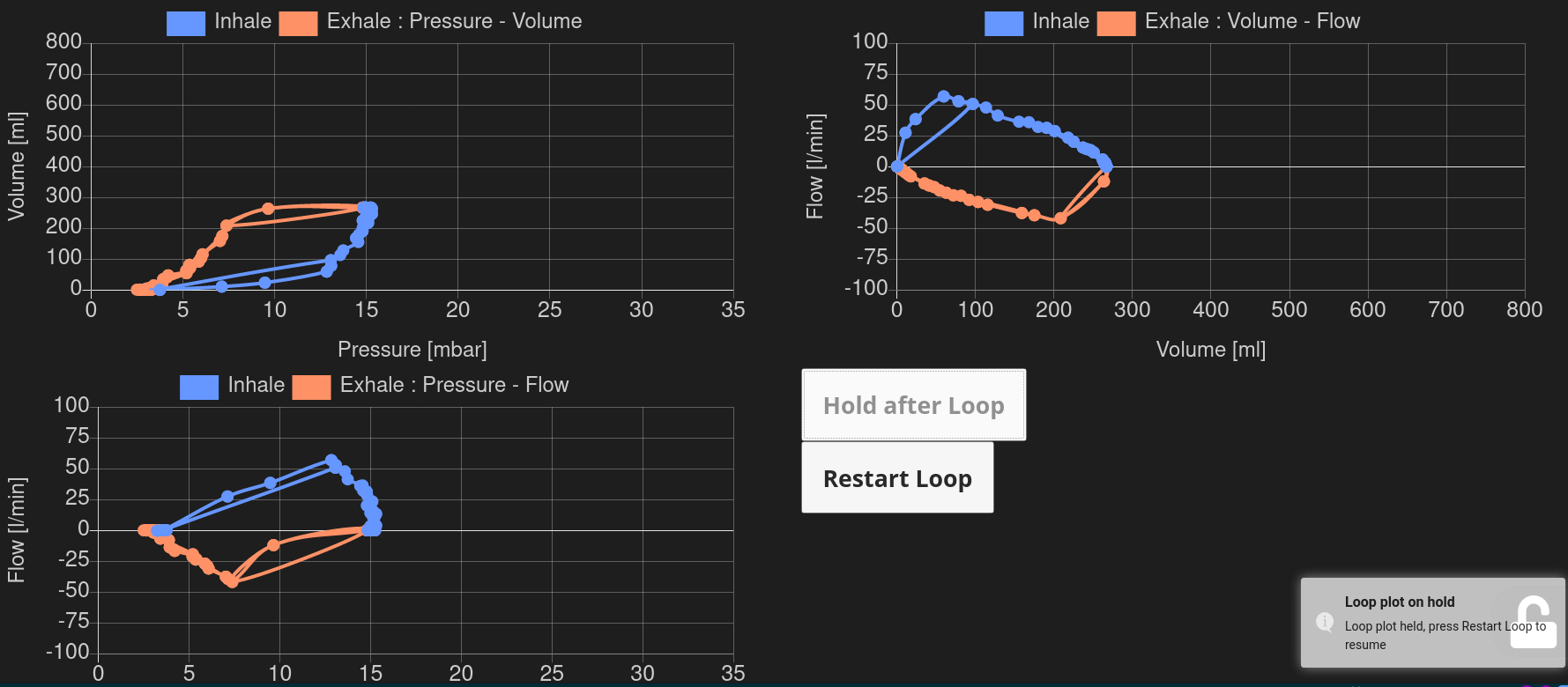}
    \caption{Example displays of the Web User Interface.  Numbers and graphs are purely indicative.}
    \label{fig:webui-homepage}
\end{figure}

The User Interface (UI) 
is designed to be familiar and readily usable in a medical environment, follows industry standards  and conventions, and respects the regulatory guidelines. Two interfaces are be provided: a Native User interface and a Web User Interface.  The Native User Interface runs on the touchscreen integrated into the ventilator unit.  It continuously displays settings, measurements, patient waveforms and accepts instructions from the clinician.  The displays are implemented in large clear fonts that can be seen easily from a distance of about 2~m away. Furthermore, the buttons or control elements are spaced well enough apart to accommodate thumb sized control. Figure \ref{fig:nativeui-homepage} contains examples of the Native UI display.  
Remote access is provided via the Web UI, accessible via WiFi or Ethernet connection to the ventilator on computer screens or mobile devices.  In this way the data of one or more patients can be collected and displayed at the nurses' station.  This also opens up the possibility of remote consulting, which can be very useful for training or patient management in remote settings.  The web interface can be configured so that full control, partial control, or no control is possible remotely. Access rights will be configurable as required. A Web UI example can be found in figure \ref{fig:webui-homepage}) 
The Native UI is automatically displayed on power up of the device.  At start-up, a mode selection screen allows for selection of one of the modes described in section \ref{subsec::VentiliationModes}.  Associated with
each mode are the set parameters to be chosen for that mode.
The parameter settings can be revisited at any time during operation.
Each setting change requires a confirmation from the user.
The UIs are designed with safety in mind.  
A locking feature of the Native UI prevents accidental touches.
Inactivity will activate automatic locking.  Locking can be
enabled/disabled by holding the lock button for 3 seconds.  Visual
feedback is given during the locking/unlocking procedure.
Indicators for the power source (mains or battery) are given on
the UI, including residual charge indicators (e.g. charge $<85\%$,
30 mins remaining).

A ``homepage'' is shown for the current ventilation mode with the most important settings, parameters, waveforms and control buttons.  Similar  to the locking feature, the Native UI defaults to presenting the homepage after a period of
inactivity by the user.  A selection is provided for time ranges
of the waveforms (i.e., showing the last 5, 15, 30 or 60 seconds).
Historical data are recorded for up to 10 days.  Encryption will
be an option for stored patient data.  Settings (or target values) and
measured values are clearly distinguishable in the UI.  The HEV Graphical User Interface is organised to respect all of the requirements listed in~\cite{mhra,who_specs,fda_specs} on both implementations, and includes the following items on the homepage:
\begin{itemize}
    \item Ventilation mode.
    \item Working inhalation pressure setting.
    \item Respiratory rate setting.
    \item Inhalation time setting.
    \item FIO$_2$ setting.
    \item Monitored plateau pressure. 
    \item Monitored respiratory rate. 
    \item Monitored PEEP.
    \item Monitored FIO$_2$. 
    \item Monitored exhaled tidal volume. 
    \item Monitored exhaled minute volume. 
\end{itemize}

The following items are plotted as a continuous graphical display, with an option given to pause the display should the clinician wish to study the waveforms more closely.
\begin{itemize}
    \item Volume vs time.
    \item Pressure vs time.
    \item Flow vs time.
\end{itemize}

Alarms are displayed in an intuitive way at the top of the screen at all times.
A dedicated alarm page gives more detail on the alarms.  An ordered
list of the last ten alarms is shown, ordered by alarm priority, and current and historical alarms are easily distinguishable. It is possible to reset or silence alarms (for a period of time).  The user
interfaces match the ``traffic light'' lamps in terms of on-screen
visualisation.  


Finally, more technical details of the internal operation and calibration
of the ventilator are provided on a separate ``expert'' page.
A visual warning indicates that this is not for normal usage.   The aim of the prototyping was to put in place all underlying software flexibility to be able to freely implement the desired UI.  Before final manufacture a full useability study will be performed in order to optimise the UI.

\section{HEV Prototypes}
\label{sec:prototypes}

\begin{figure}[h]
\begin{center}
  \includegraphics[width=0.95\linewidth]{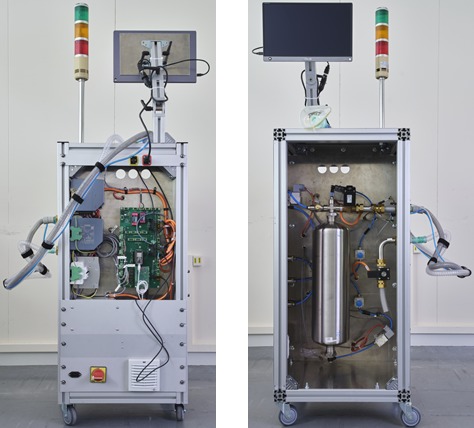}
  \caption{HEV prototypes as tested for full functionality.}
  \label{fig:HEVprototype}
  \end{center}
\end{figure}

Three prototypes were constructed with an identical mechanical design in order to allow work in parallel on different aspects of the design implementation.  
%
The prototypes (figure~\ref{fig:HEVprototype}) are designed as standalone units, each with two separate compartments, pneumatic and electrical, separated by a support plate.  The mechanical description below as well as the functional tests described were all performed on  these prototypes.

\begin{figure}[h]
\begin{center}
  \includegraphics[width=0.95\linewidth]{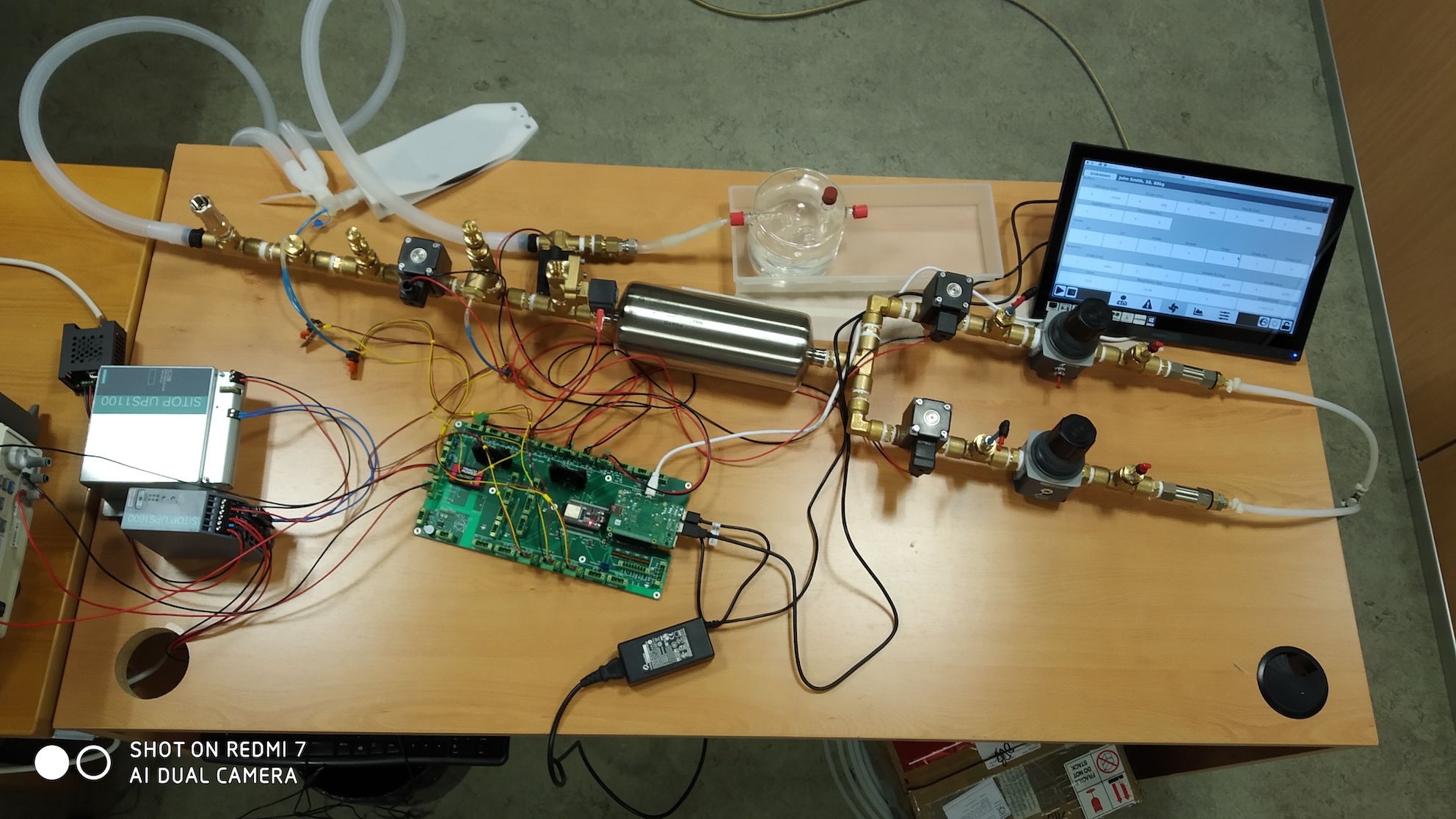}
  \caption{Functional table top prototype produced for collaboration and development purposes.}
  \label{fig:dimitri}
  \end{center}
\end{figure}

In parallel, a fourth prototype, illustrated in figure~\ref{fig:dimitri}, was developed at the Galician Institute of High Energy Physics at the University of Santiago de Compostela (IGFAE/USC) with a two-fold goal. Firstly, we wished to demonstrate that such a respirator could be developed easily in other places than CERN.  Secondly, having successfully reproduced the design within a short timescale, we were able to launch additional development and contributions to the control software. In addition, this triggered discussions with local hospitals and physicians, further improving the design of the device. 

\subsection{Prototype mechanical design}
\label{sec::mechanical}



The prototype mechanical design was created with the aim of enabling the production of a fully specified and functional ventilator.  Additional space has been deliberately added, so that the working prototypes could be built as quickly as possible, and to ensure that there will be no problem to interchange components for others with similar properties if this is necessary for full medical compatibility.  The resulting design iis a unit with approximate dimensions $500 \times 500 \times 1350$~mm. Note that due to this choice, the machine is larger than necessary and not in its optimal packaging; this can be easily optimised at the manufacturing stage.  

The simple mechanical design was chosen to feature standard components, not commonly used in respiratory equipment. This would enable it to be produced locally by adapting the parts to the availability of local suppliers. The ventilator is simple to construct and requires little or no machining. A list of the typical parts required includes: 


%
\begin{itemize}
    \item A proportional solenoid inhale valve with 8 mm internal diameter.
    \item Two fast acting solenoid valves for inlet of air and oxygen into the buffer volume. 
    \item A solenoid valve for purging. It could be low volume, that in case of over-pressure only some air is released.
    \item An exhale ON/OFF valve with large diameter internal orifice.
    \item Five check valves.
    \item Two pressure relief valves, one for the patient circuit and one for the buffer.
    \item A 10 L buffer container. 
    \item Two pressure regulators, able to operate up to a maximum pressure of 10 bar,
    one for oxygen and one for air. 
    \item Eight pressure sensors.
    \item A bidirectional differential pressure sensor for the flow measurement.
    \item Two temperature sensors.
\end{itemize}

In addition to these parts, there is a list of required and optional accessories, described in section~\ref{sec::turbineaccessories}, such as the PEEP valve, breathing circuit, filters and humidifiers.

Functionally, the prototype designs follow the concept described above.  The resulting cabinet is mounted on wheels, can easily be moved by one person, is very stable, and provides a convenient surface to mount the display at head height.  The cabinet is closed with doors, so that easy access for cleaning is possible.  The cabinet is subdivided internally into two separate compartments, front and back, housing the pneumatic and electronics components separately, which provides protection against explosion risk from potential oxygen leaks.  The air tubes connect through a standard bulkhead thread connector on the outside.  In this way it is easily replaceable to match hospital connection standards around the world.

\begin{figure}[h]
\begin{center}
  \includegraphics[width=1.0\linewidth]{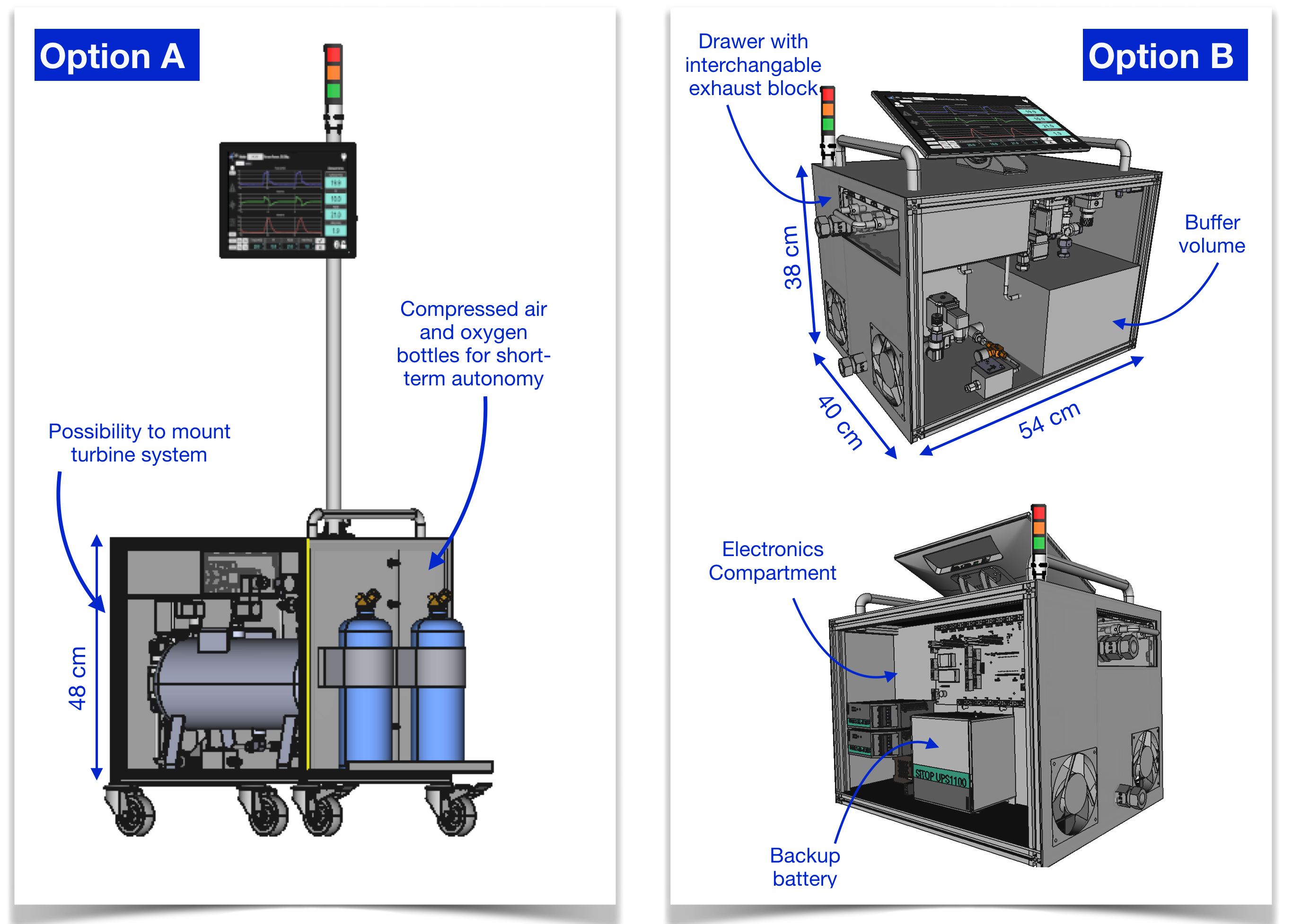}
  \caption{Potential alternative mechanical designs with an identical functionality to the HEV prototypes.}
  \label{fig:compact_design}
  \end{center}
\end{figure}

\subsection{Alternative mechanical design}

The prototypes have been built with a deliberately large amount of space to allow rapid development and exchange of parts. The final ergonomics of the HEV may look quite different depending on the requirements in the region of deployment and the accessories included. The HEV collaboration has provided two different mechanical designs to which the HEV design could be adapted to fulfill different needs, which are illustrated in figure~\ref{fig:compact_design}.  Option A is more compact and can be mounted on wheels or a trolley.  Space is provided to support oxygen and compressed air bottles, as well as the turbine system, such that the entire system and accessories can be provided as one integrated unit, which can be desirable for certain geographical locations.  Option B is a still more compact and light version, for which the total dimensions are comparable to existing commercial ventilators and the weight is targetted to be around 25 kg.  The touch screen can be folded away for transport and the ventilator easily mounted on a trolley.  Both options are identical in functionality to the HEV prototypes which have been built and tested.

\section{Prototype test results}
\label{sec:testresults}
\subsection{Setup}

\begin{figure}[!h]
\begin{center}
  \includegraphics[width=0.95\linewidth]{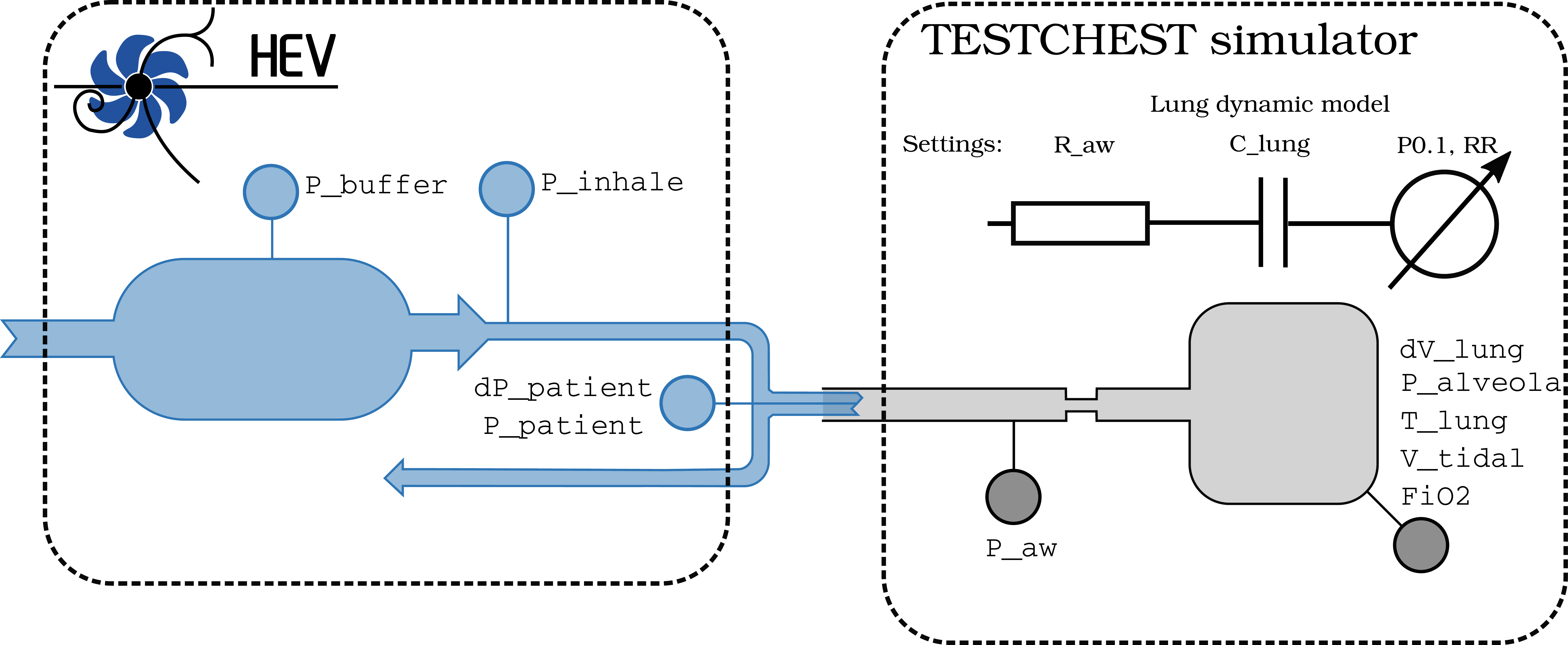}
  \caption{Test setup schematics.}
  \label{fig:HEVSetup}
  \end{center}
\end{figure}

The HEV prototype is tested by connecting it to a lung simulator through a coaxial breathing circuit set, as shown in figure~\ref{fig:HEVSetup}. 
The coaxial  breathing circuit set\footnote{\href{https://www.hamilton-medical.com/Products/Accessories-and-Consumables/E-catalog/Detail~260128~Kit-de-circuit-respiratoire--coaxial~b4327ab7-0199-4dab-a095-3fd6a1831734~.html}{Hamilton PN 260128.}} has a differential pressure based flow sensor\footnote{\href{https://www.hamilton-medical.com/Products/Accessories-and-Consumables/E-catalog/Detail~281637~Capteur-de-d\%C3\%A9bit~7e01c9fd-4af9-48ed-a6bf-89f5beed4771~.html}{Hamilton PN 281637.}} whose readout is embedded in HEV (\dPpatient).   Alternatives to this sensor are available with other manufacturers~\footnote{\href{https://www.intersurgical.com/products/critical-care/transport-adult-22mm-breathing-systems}{Intersurgical P/N 2072000}}, and a specific breathing circut for HEV in order to decrease reliance on the existing supply chain can also be an option.  The lung simulator is a TestChest light~\footnote{Organis GMBH, Landquart, www.organis-gmbh.com}\cite{testchest}, which allows for the change the mechanical parameters of the lung (resistance of the airways, compliance of the lung) as well as generates adjustable spontaneous breathing. 
The setup is equivalent to the MHRA~\cite{mhra} and ISO 80601-2-12 figure 201.102~\cite{iso80601-2-12:2020}. 

\subsection{Target Pressure Performance}

\begin{figure}[!ht]
    \centering
    \includegraphics[width=1.\textwidth]{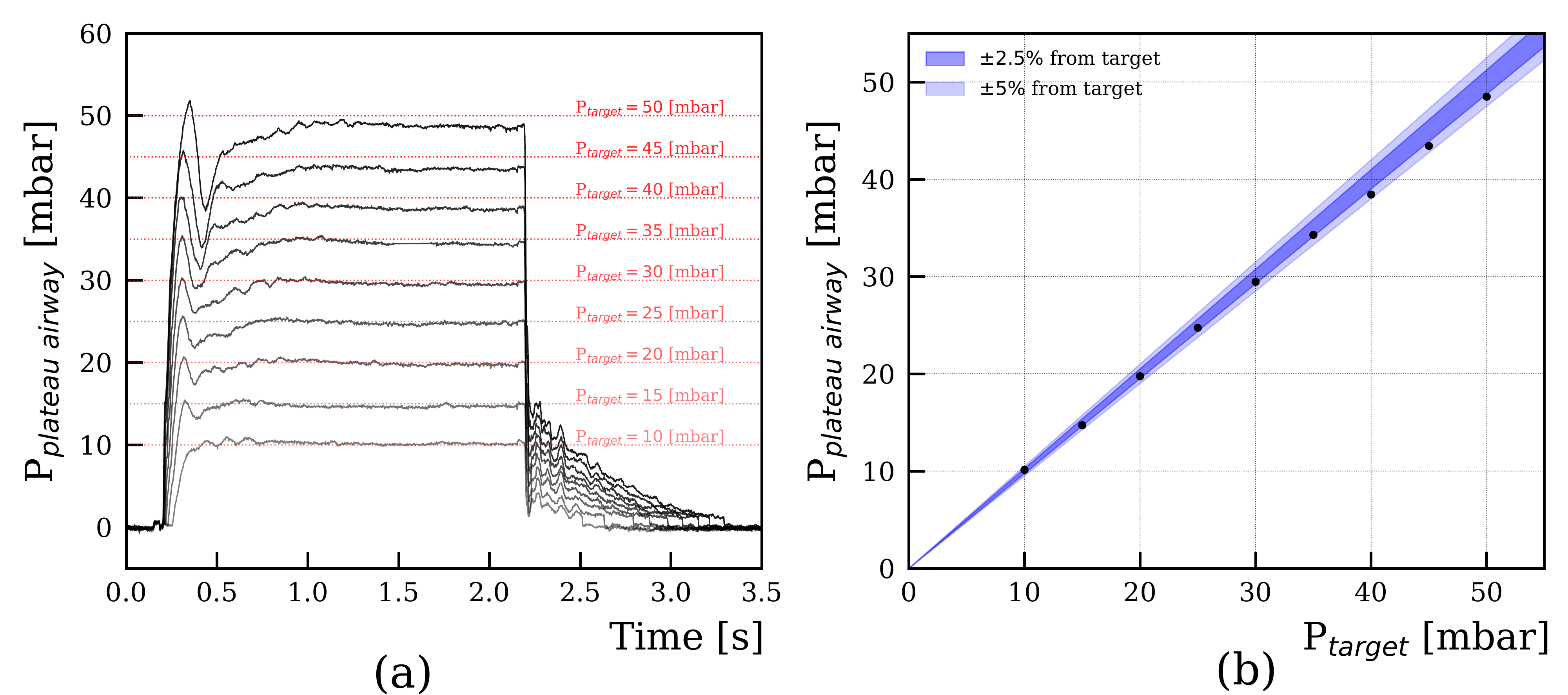}
    \caption{Left: pressurisation of a 50~ml/mbar compliance, 5~mbar/l/sec resistance patient with various target pressures. The inhalation time is set to 1.5 seconds, and is followed by a pause of 0.5 seconds. Right: the deviation of the inhale pressure from the target is computed during the pause.}
    \label{fig:ptarget}
\end{figure}

As discussed in section~\ref{sec::conceptualdesign::pneumatic}, when the inhale starts, the proportional valve \valveinhale opens in a controlled way in order to reach the target pressure in a given time. This is done through a PID controller that monitors the \texttt{P\_{inhale}} pressure measurement as input for the inhale valve opening control. As illustrated in figure~\ref{fig:ptarget}, any pressure level within the setting range can be reached and the system stays stably locked to this value with an uncertainty below 3\% in most of the cases.

Further tests will be performed in presence of leaks, but even with the highest leak setting of the TestChest, no difference in the pressure profile is visible.

\begin{figure}[!hb]
    \centering
    \includegraphics[width=.7\textwidth]{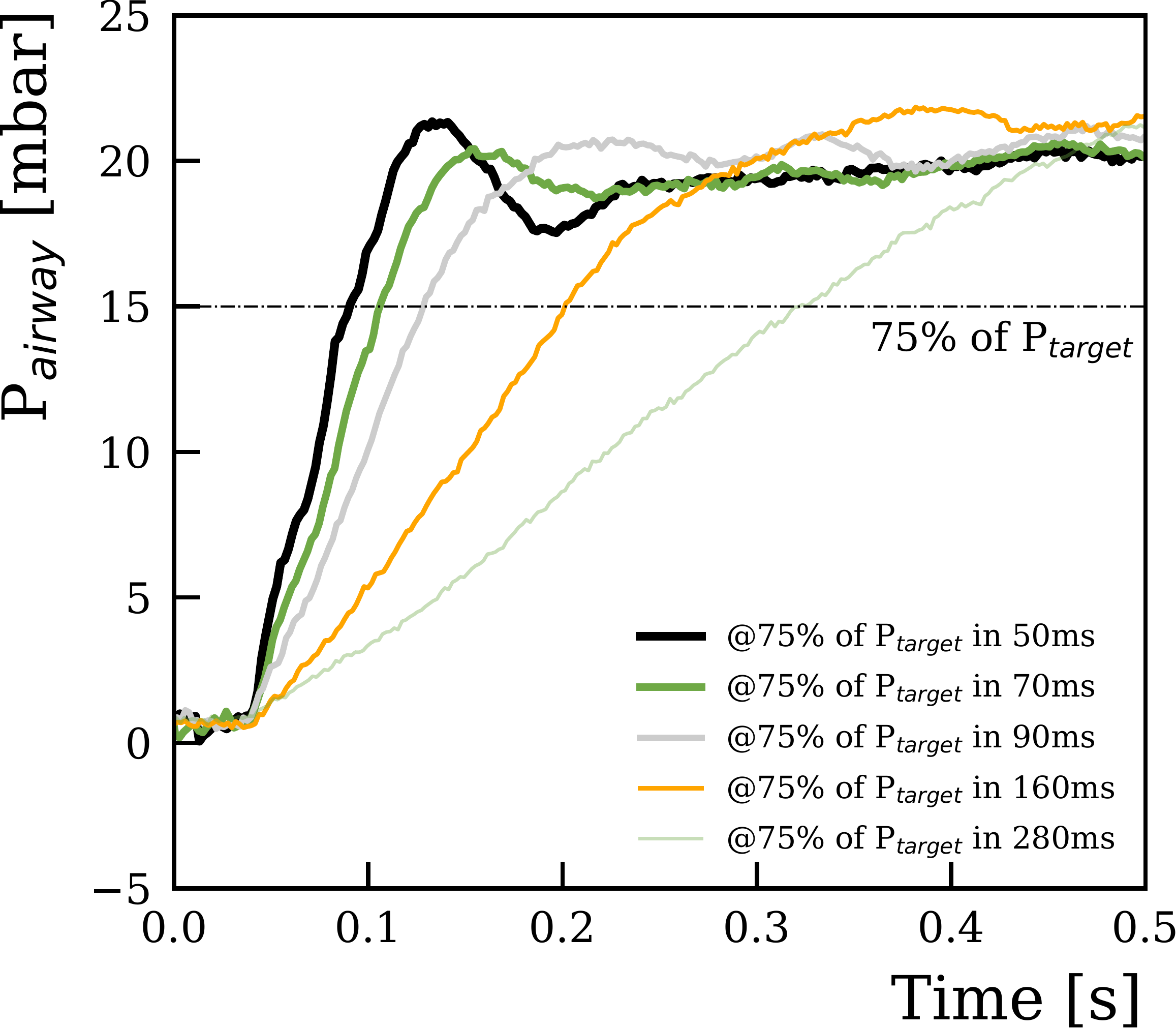}
    \caption{Rising edge of the inhalation for various rise time setting. The lung compliance here is 50~ml/mbar for a resistance of 5~mbar/l/s}
    \label{fig:pcacrisetime}
\end{figure}

The time needed to reach the maximum of pressurisation can be tuned between the fastest setting of about 50~ms and about 300~ms, as illustrated in figure~\ref{fig:pcacrisetime}. This gives the clinician the flexibility to use the fastest response of the ventilator, typically for intubated, sedated patients, through to slower rise times which may be useful at other periods of recovery.

\begin{figure}[!h]
    \centering
    \includegraphics[width=.94\textwidth]{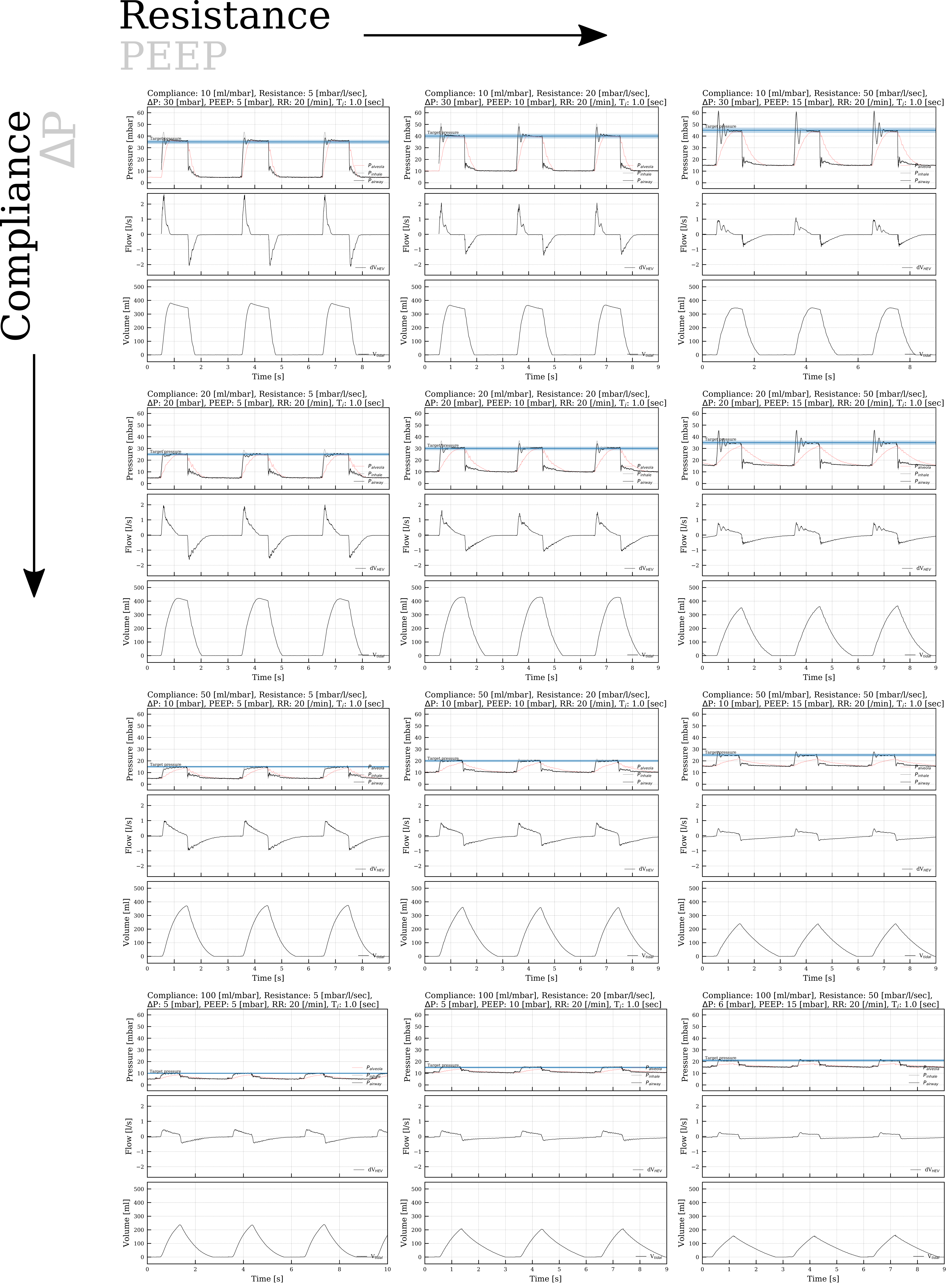}
    \caption{Pressure, flow and volume registered for patient configurations with increasing lung compliance and airway resistance. As expected, the flow is more peaked with reduced resistance and compliance.}
    \label{fig:pcac}
\end{figure}

In figures~\ref{fig:pcac}, the response of the HEV ventilator to patients with compliance varying from 10 to 100 ml/mbar and with resistance from 5 to 50 mbar/l/s is shown.  The PEEP values range from 5 to 15 \cmWater and target pressure up to 45\cmWater are tested. The behaviour of the ventilator is illustrated in figure~\ref{fig:pcac}.

When the patients shows no airway resistance or when their lung compliance is low enough, the flow can fully develop during the allocated inhalation time such that the tidal volume can be computed from the product of the compliance and the differential pressure.
The slowing down of the flow with the increased resistance and increased compliance is also well visible. Lower compliance patients could not be tested with this lung simulator but will be the subject of a dedicated test later; the performance is expected to be good.

\clearpage

\subsection{Inhale Trigger Peformance}

The inhale and exhale trigger are essential to guarantee the comfort of the patients and to ensure fast recovery time. The trigger functionalities were developed with this aspect in mind and particular effort was made to qualify them.

\begin{figure}[!h]
    \centering
    \includegraphics[width=0.4\textwidth]{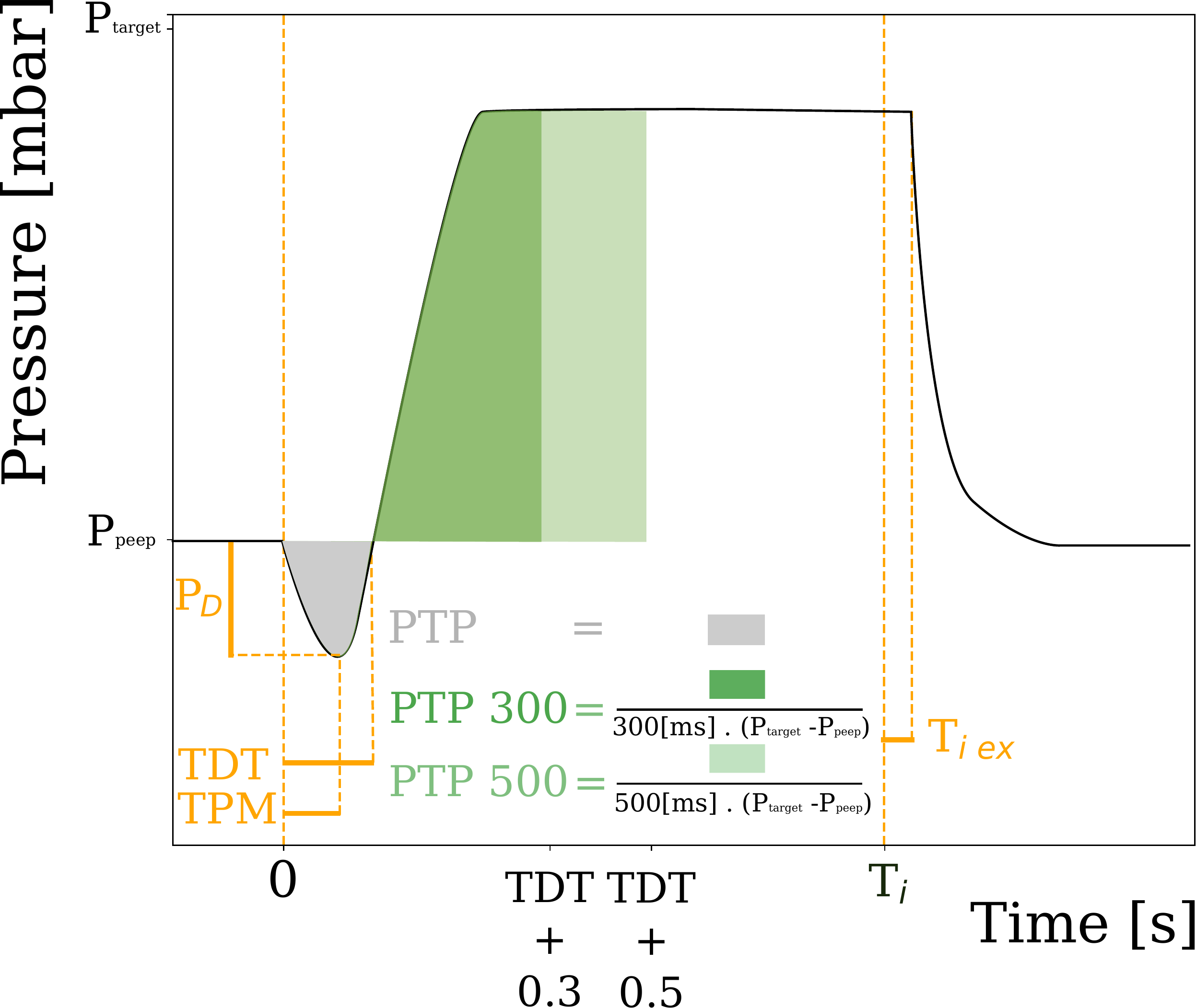}
    \caption{Measured parameters to qualify the inhalation trigger. The variables used in this work are based on the work presented in \cite{delgado2017} and \cite{vignaux2007}}
    \label{fig:ptp300}
\end{figure}

\paragraph*{Inhale trigger algorithm description} 
Whenever the patient initiates a breath, the proximal flow sensor sees an increased flow and an under-pressure. The increase in flow is used as a trigger for the inhale sequence. The inhale trigger algorithm works in the following way. Whenever the flow reaches 10\% of the maximal exhale flow, the window allowing for a inhale trigger is opened. This condition is by definition always met when the patient initiates an inhalation. The expected flow at a given instant is computed by a linear regression from a given time window before that point. It provides the baseline, to which the measured flow is compared. If the measured flow, corrected for the baseline, is above a threshold that can range from 0.2 l/min to 20 l/min, then the inhalation starts. Lower thresholds are sensitive to noise, in particular that induced by the heart-lung interactions. This threshold is set by the clinician.


\paragraph*{Inhale trigger qualification}
To qualify the performance of the inhale trigger, the variables defined in~\cite{delgado2017} are used. Figure~\ref{fig:ptp300} illustrates those variables: the Time to Minimum Pressure, TPM, is defined as the time between the beginning of the inhalation effort and the minimum value measured reached by \Ppatient, and the Trigger Delay Time TDT is defined as the time between the beginning of the inhalation effort and the moment the pressure returns to zero. Values of TPM in commercial ventilators typically vary between 50 and 150~ms depending on the inhalation effort, see for example~\cite{delgado2017},  while TDT, which should ideally be below 150~ms so as not to be felt by the patient, varies for commercial ventilators in practice between 90 and 250 ms~\cite{delgado2017}.
The pressure-time product during trigger, PTP, is represented in figure~\ref{fig:ptp300} by the grey area and represents the effort until the pressure is effective. It ranges from 0.02 to 0.3 mbar.s in commercial ventilator~\cite{delgado2017}. The ideal PTP300 percentage and ideal PTP500 percentage,referred to as PTP300 and PTP500 in the following, are respectively the ratio of the pressure integral over the 300~ms (500~ms) following the trigger delay (the area of the green regions in figure~\ref{fig:ptp300}) and the ideal PTP at 300 and 500~ms. It should be as large as possible, with typical commercial ventilators exhibiting values between 10 and 50\% (20 and 75\%) for PTP300 (PTP500) depending on the inhalation effort~\cite{delgado2017}.

To test the inhalation trigger, the same lung parameters, pressurisation parameters and inhalatory effort than in~\cite{delgado2017} were used. TestChest was set with a compliance of 50 mbar/ml and a resistance of 5 mbar/l/s, the breath cycles at 12 respiration per minutes consisted in a 1 s inspiration with a constant inspiratory flow giving a occlusion pressure at 100~ms (P$_{0.1}$) of 2\cmWater (low effort) and 4\cmWater (high effort) and several pressurisation parameter: $\Delta$P of 10,15 and 20 mbar with a PEEP of 0 and 5 mbar. The inhale trigger threshold is set to 0.5 L/min. 

Table~\ref{tab:trig} summarises the results of the inhalation trigger qualification. Comparing the results to the ventilator studied in ~\cite{delgado2017}, HEV inhale trigger appears to perform very well.

\begin{table}[]
\begin{tabular}{l|llllll|llllll|}
P$_{0.1}$ [mbar] & \multicolumn{6}{c|}{2} & \multicolumn{6}{c|}{4} \\ \cline{2-13} 
PEEP [mbar] & \multicolumn{3}{c|}{0} & \multicolumn{3}{c|}{5} & \multicolumn{3}{c|}{0} & \multicolumn{3}{c|}{5} \\ \cline{2-13} 
$\Delta P$  [mbar] & \multicolumn{1}{c|}{10} & \multicolumn{1}{c|}{15} & \multicolumn{1}{c|}{20} & \multicolumn{1}{c|}{10} & \multicolumn{1}{c|}{15} & \multicolumn{1}{c|}{20} & \multicolumn{1}{c|}{10} & \multicolumn{1}{c|}{15} & \multicolumn{1}{c|}{20} & \multicolumn{1}{c|}{10} & \multicolumn{1}{c|}{15} & \multicolumn{1}{c|}{20} \\ \hline
TPM {[}ms{]} & 100 & 99 & 79 & 121 & 104 & 74 & 118 & 69 & 74 & 95 & 95 & 72 \\
TDT {[}ms{]} & 130 & 105 & 101 & 139 & 112 & 98 & 150 & 105 & 98 & 135 & 107 & 96 \\
PD [mbar] & 2 & 1.9 & 1.6 & 1.9 & 2 & 1.7 & 4.7 & 3.7 & 4 & 4.1 & 3.9 & 2.6 \\
PTP [mbar] & 0.09 & 0.03 & 0.02 & 0.12 & 0.08 & 0.07 & 0.24 & 0.15 & 0.13 & 0.28 & 0.2 & 0.18 \\
PTP300 {[}\%{]} & 28.8 & 41.4 & 43.8 & 29.1 & 38.6 & 46.4 & 24.8 & 36.6 & 41.0 & 29.8 & 36.1 & 43.5 \\
PTP500 {[}\%{]} & 39 & 53.8 & 55.6 & 43.7 & 51.2 & 58.8 & 34.6 & 44.6 & 49.1 & 38.2 & 44. & 52.2
\end{tabular}
\caption{Results of the inhale trigger characterisation.}
\label{tab:trig}
\end{table}

The trigger response is studied again in presence of leaks. Only one set of pressurisation is used, with PEEP at 0\cmWater and $\Delta P$=20\cmWater. Four settings are tried: no leaks, weak, medium and strong leaks as set by TestChest, and results are reported in table~\ref{tab:leak}.  No significant difference is observed.

\begin{table}[]
\begin{tabular}{l|llll}
 & \multicolumn{1}{c}{No leaks} & \multicolumn{1}{c}{Small} & \multicolumn{1}{c}{Medium} & \multicolumn{1}{c}{Large} \\ \hline
TPM {[}ms{]} & 74 & 72 & 70 & 70 \\
TDT {[}ms{]} & 98 & 94 & 94 & 94 \\
PD {[}mbar{]} & 4 & 4 & 3.7 & 3.4 \\
PTP {[}mbar.s{]} & 0.13 & 0.13 & 0.11 & 0.1 \\
PTP300 {[}\%{]} & 41 & 40.5 & 41.4 & 41.1 \\
PTP500 {[}\%{]} & 49.1 & 48.5 & 49.3 & 49.1
\end{tabular}
\caption{Inhale trigger sensitivity to different leakage level for lungs with compliance of 50 mbar/ml and resistance of 5 mbar/l/s}
\label{tab:leak}
\end{table}
\paragraph*{Exhale trigger algorithm description} The implementation of the exhale trigger is more straightforward. When the inhale flow decreases down to a fraction of the maximum inhale flow, the exhale phase is triggered.

\paragraph*{Exhale trigger qualification}
In order to qualify the performance of the exhale trigger, $\mathrm{T_{I_{ex}}}$ defined as the duration of pressurisation by the ventilator in excess with respect to $\mathrm{T_I}$, which represents the true duration of inhalation by the patient, is measured as in~\cite{vignaux2007}. This is illustrated in figure~\ref{fig:ptp300}. It is not a property of the ventilator per se, but by appropriate tuning of the exhalation trigger it should be possible to bring it to below 100~ms. Values of $\mathrm{T_{I_{ex}}}$ below 10~ms are achieved.

\subsection{Oxygen mixing test}

Because the mixing is performed in the buffer in a phase which is physically uncorrelated to the patient the breath cycle (i.e. during patient exhalation when the buffer is disconnected from the patient), we perform the mixing test independently from the test of the ventilator modes.

The O$_2$ concentration in the buffer is controlled by changing the relative opening time of \valveoin and \valveairin.  For a given O$_2$ concentration setting, the opening times can be computed. After stabilisation of the measured O$_2$ concentration (\FIOxygen) in the lung simulator, the \FIOxygen is compared to the set value. Further control will be introduced in the future by regulating the opening time from a feedback of the measured O$_2$ concentration in the buffer.

\begin{figure}[h]
    \centering
    \includegraphics[width=0.4\textwidth]{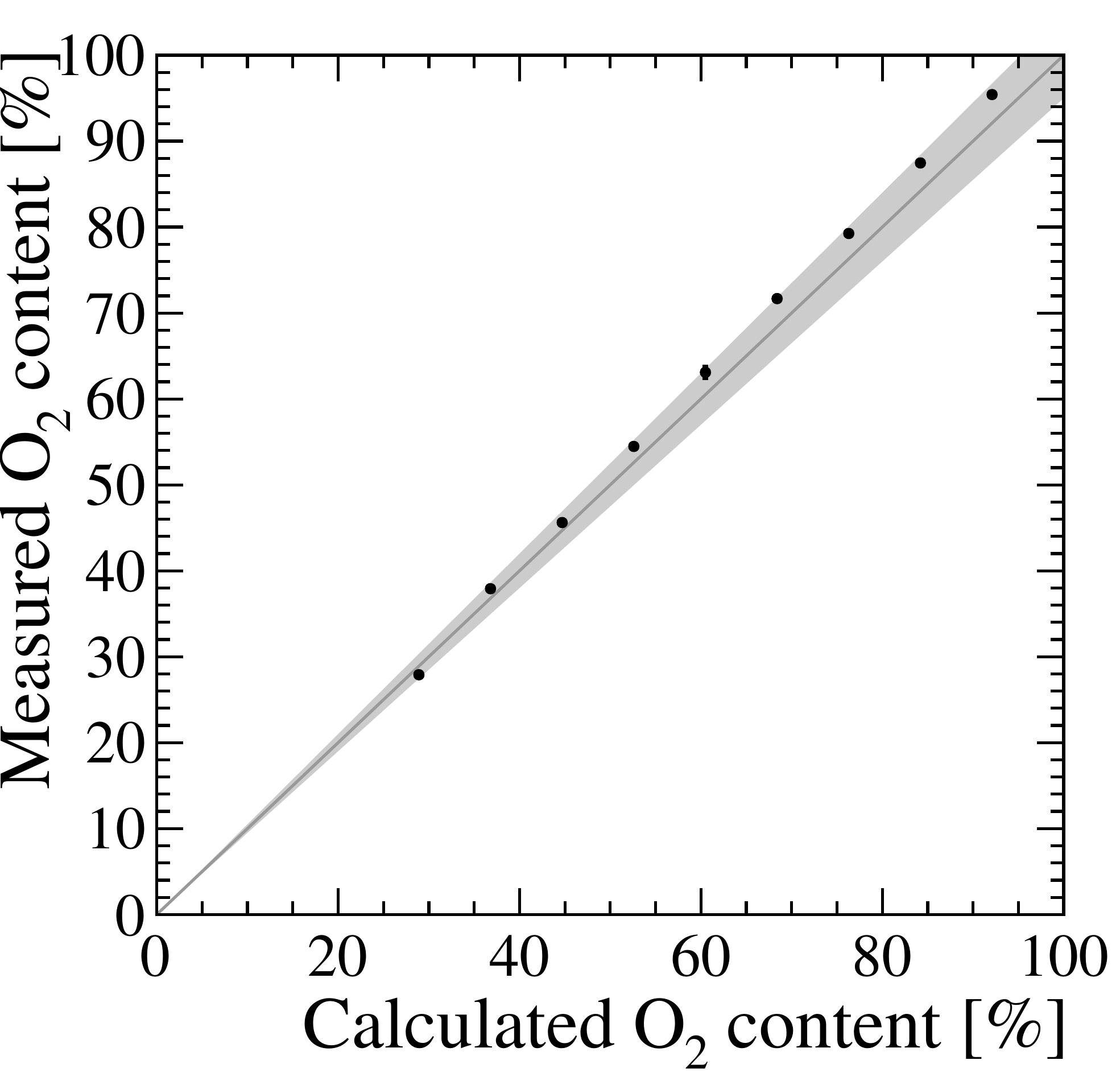}
    \caption{Measured \FIOxygen as function of the expected O$_2$ percentage as calculated from the relative time opening of \valveoin and \valveairin. The grey band represents the region within $\pm5\%$ of the set value. }
    \label{fig:oxygen}
\end{figure}

The measured \FIOxygen as function of the expected O$_2$ percentage as calculated from the relative time opening of \valveoin and \valveairin is shown in figure~\ref{fig:oxygen}. The measured \FIOxygen in the lung is within 5\% of the set value, which is an acceptable performance. Further tuning to the valve opening time can be done in order to correct for the small non-linearity in the response.

\pagebreak




\clearpage
\section{Conclusion}
\label{sec:Conclusion}

HEV has been developed to be a high-quality, low-cost ventilator, suitable for use in a hospital setting.  The design is intended for easy and fast manufacturing that can be performed in a decentralised way with affordable and readily available parts.  The central concept of the design with a gas accumulator gives many advantages in terms of robustness, safety, affordability and precise ventilation behaviour.  The  electrical design is conceived in a modular way for quick prototyping and deployment, which facilitates mass production. The design is intended to be robust and adaptable for a wide range of geographical deployment, including in regions where compressed air may not be readily available and a turbine alternative can be used. Three prototypes have been manufactured and have been tested in situ under clinical supervision with the full range of simulated patients defined in the MHRA specifications and the results are presented in this paper.  HEV has also been tested at the ETH Zurich Chair of Product Development and Engineering Design Ventilator test rig.  In pressure control mode HEV accurately achieves the target pressures, with fast rise time which is tuneable to slower times on clinician request. Special attention has been paid to the inhale and exhale triggers to optimise patient comfort.  The inhale trigger, based on the flow measurement, accurately reacts to the patient effort, with short rise times and excellent PTP values.  The system displays and monitoring use concepts familiar to particle physics such as the possibility for remote monitoring from screens or mobile devices, data logging for quality control and performance monitoring, and remote training.

As far as production is concerned, it is foreseen, on the one hand, to enable this through providing partner academic institutions with the detailed design for these institutions to follow up in accordance with local possibilities and standards;  on the other hand, directly through industry, non-governmental, governmental and international organizations, such as the World Health Organisation (WHO), for which purpose discussions are ongoing and contacts have been established with potential partners.
Every effort is being made to finalise the design of the HEV in accordance with the state-of-the-art best practices and standards, but the formal certification process should of course be initiated by the parties that decide to place this device on the market.  The hardware and software design has been done in a flexible way which allows the development of different modes of operation, for instance volume control modes which in principle can be developed and applied as a firmware update.  In addition, the HEV prototypes can be used as a testbench to quickly implement and test novel algorithms or hardware updates, and in this way could provide a fresh avenue for medical research.

\section*{Acknowledgements}

\noindent  We express our gratitude to
Giovanni Anelli, Amy Bilton, Paola Catapano, Manuela Cirilli, Paolo Chiggiato, Beniamino Di Girolamo, Doris Forkel-Wirth, Benjamin Frisch, Samuel Herzog, Lucie Pocha, Javier Serrano, Erik Van Der Bij, Jens Vigen and Maarten Wilbers from CERN, Geneva, Switzerland, the CERN against COVID-19 task force, and to Simon Cohen from Monash Children's Hospital, Melbourne, Australia, for many illuminating discussions and much practical support.

\addcontentsline{toc}{section}{References}
\printbibliography

\newpage

\centerline{\large\normalfont\bfseries HEV collaboration}
\begin{flushleft}
\small
J.~Buytaert$^{1,*}$, 
A.~Abed~Abud$^{1,2}$,
P.~Allport$^{26}$, 
A. Pazos \'Alvarez$^{11}$,
K.~Akiba$^{3}$,
O.~Augusto~de~Aguiar Francisco$^{1,4}$, 
A.~Bay$^{10}$, 
F.~Bernard$^{10}$, 
S.~Baron$^{1}$,
C.~Bertella$^{1}$,
J.~Brunner$^{21}$,
T.~Bowcock$^{2}$, 
M.~Buytaert-De Jode, 
W.~Byczynski$^{1,8}$,
R.~De~Carvalho$^{23}$,
V.~Coco$^{1}$, 
P.~Collins$^{1,*}$,
R.~Collins$^{18}$,
N.~Dikic$^{1}$, 
N.~Dousse$^{23}$,
B.~Dowd$^{16}$,
R.~Dumps$^{1}$, 
P.~Durante$^{1}$, 
W.~Fadel$^{1}$,
S.~Farry$^{2}$,
A.~Fern\'andez Prieto$^{11}$,
G.~Flynn$^{16,25}$, 
V.~Franco Lima$^{2}$,
R. Frei$^{10}$, 
A.~Gallas Torreira$^{11}$,
R.~Guida$^{1}$, 
K.~Hennessy$^{2}$,
A.~Henriques$^{1}$,
D.~Hutchcroft$^{2}$, 
S.~Ilic$^{7}$, 
A.~Jevtic$^{7}$, 
C.~Joram$^{1}$, 
K.~Kapusniak$^{1}$,
E.~Lemos~Cid$^{11}$,
J.~Lindner$^{9}$, 
R.~Lindner$^{1}$,
M.~Milovanovic$^{1,6}$,
S.~Mico$^{1}$,
J.~Morant$^{1}$,
M.~Morel$^{1}$,
G.~M\"{a}nnel$^{20}$,
D.~Murray$^{4}$, 
I.~Nasteva$^{5}$, 
N.~Neufeld$^{1}$, 
I.~Neuhold$^{1}$,
F.~Pardo-Sobrino L\'opez$^{19}$,
E.~P\'erez Trigo$^{11}$,
G.~Pichel Jallas,
E.~Pilorz$^{1}$,
L.~Piquilloud$^{17}$,
X.~Pons$^{1}$, 
D.~Reiner$^{13}$, 
C.~Roosens$^{24}$,
P.~Rostalski$^{20}$,
B.~Schmidt$^{1}$,
E.~Saucet$^{23}$, 
F.~Sanders$^{1}$, 
C.~Sigaud$^{1}$,
B.~Schmidt$^{1}$, 
P.~Schoettker$^{17}$
R.~Schwemmer$^{1}$,
H.~Schindler$^{1}$,
A.~Sharma$^{1}$,
P.~Svihra$^{4}$,
J.~van~Leemput$^{24}$,
L.~Vignaux$^{22}$,
F.~Vasey$^{1}$,
H.~Woonton$^{14,15}$,
K.~Wyllie$^{1}$.
\bigskip\newline{\it
\footnotesize


$  ^{1}$European Organization for Nuclear Research (CERN), Geneva, Switzerland\\
$ ^{2}$Oliver Lodge Laboratory, University of Liverpool, Liverpool, United Kingdom\\
$ ^{3}$Nikhef National Institute for Subatomic Physics, Amsterdam, Netherlands\\
$ ^{4}$Department of Physics and Astronomy, University of Manchester, Manchester, United Kingdom\\
$ ^{5}$Universidade Federal do Rio de Janeiro (UFRJ), Rio de Janeiro, Brazil\\
$ ^{6}$Deutsches Elektronen-Synchrotron (DESY), Platanenallee 6, 15738 Zeuthen, Germany \\
$ ^{7}$University of Ni\v{s}, Ni\v{s}, Serbia \\
$ ^{8}$Tadeusz Kosciuszko Cracow University of Technology, Cracow, Poland \\
$ ^{9}$University of Applied Sciences Offenburg, Offenburg, Baden-Wuerttemberg,  Germany \\
$ ^{10}$Institute of Physics, Ecole Polytechnique F\'ed\'erale de Lausanne (EPFL), Lausanne, Switzerland \\
$ ^{11}$Instituto Galego de F\'{i}sica de Altas Enerx\'{i}as (IGFAE), Universidade de Santiago de Compostela, Santiago de Compostela, Spain \\
$ ^{13}$John Curtin School of Medical Research, Canberra, Australia\\
$ ^{14}$Monash Health, Melbourne, Australia\\
$ ^{15}$Dandenong Hospital, Melbourne, Australia\\
$ ^{16}$Prince of Wales Hospital, New South Wales, Australia\\
$ ^{17}$Centre Hospitalier Universitaire Vaudois, Lausanne, Switzerland \\
$ ^{18}$College of Veterinary Medicine, Cornell University, Ithaca, NY, USA\\
$^{19}$Anesthesiology-Reanimation and Pain Therapuetics Service,
Lucus Augusti University Hospital, Lugo, Spain\\
$^{20}$Institute for Electrical Engineering in Medicine, University of L\"{u}beck, L\"{u}beck, Germany\\
$^{21}$Neosim AG, CH-7000 Chur, Switzerland\\
$^{22}$Cardio-Respiratory Units, H\^{o}pital de La Tour, Meyrin, Switzerland\\
$^{23}$H\^{o}pitaux Universitaires de Gen\`{e}ve, Gen\`{e}ve, Switzerland\\
$^{24}$GZA hospitals, Antwerp, Belgium\\
$^{25}$University of New South Wales, Sydney, Australia\\
$^{26}$Particle Physics Group, School of Physics and Astronomy, University of Birmingham, United Kingdom\\
$^{*}$Corresponding authors: Jan.Buytaert@cern.ch and Paula.Collins@cern.ch}
\end{flushleft}

\end{document}